\documentclass[twocolumn]{aastex63}

\usepackage[version=4]{mhchem}
\graphicspath{{./}{figures/}}
\usepackage{diagbox}
\accepted{June 18, 2020}
\shorttitle{Class I 3mm chemical survey}
\shortauthors{Le Gal et al.}
\begin{document}

\title{A 3mm chemical exploration of small organics in Class I YSOs}
\correspondingauthor{Romane Le Gal}
\email{romane.le$\_$gal@cfa.harvard.edu}

\author[0000-0003-1837-3772]{Romane Le Gal}
\affiliation{Center for Astrophysics \textbar Harvard \& Smithsonian, 60 Garden St., Cambridge, MA 02138, USA}

\author[0000-0001-8798-1347]{Karin I. \"Oberg}
\affiliation{Center for Astrophysics \textbar Harvard \& Smithsonian, 60 Garden St., Cambridge, MA 02138, USA}

\author[0000-0001-6947-6072]{Jane Huang}
\affiliation{Center for Astrophysics \textbar Harvard \& Smithsonian, 60 Garden St., Cambridge, MA 02138, USA}

\author[0000-0003-1413-1776]{Charles J. Law}
\affiliation{Center for Astrophysics \textbar Harvard \& Smithsonian, 60 Garden St., Cambridge, MA 02138, USA}

\author{Fran\c cois M\'enard}
\affiliation{Universit\'e Grenoble Alpes, CNRS, IPAG, F-38000 Grenoble, France}

\author{Bertrand Lefloch}
\affiliation{Universit\'e Grenoble Alpes, CNRS, IPAG, F-38000 Grenoble, France}

\author{Charlotte Vastel}
\affiliation{IRAP, Universit\'e de Toulouse, CNRS, UPS, CNES, 31400 Toulouse, France}

\author[0000-0002-6729-3640]{Ana Lopez-Sepulcre}
\affiliation{Universit\'e Grenoble Alpes, CNRS, IPAG, F-38000 Grenoble, France}
\affiliation{IRAM, 300 rue de la piscine, F-38406 Saint-Martin d’Hères, France}

\author[0000-0002-5789-6931]{C\'ecile Favre}
\affiliation{Universit\'e Grenoble Alpes, CNRS, IPAG, F-38000 Grenoble, France}

\author[0000-0001-9249-7082]{Eleonora Bianchi}
\affiliation{Universit\'e Grenoble Alpes, CNRS, IPAG, F-38000 Grenoble, France}

\author[0000-0001-9664-6292]{Cecilia Ceccarelli}
\affiliation{Universit\'e Grenoble Alpes, CNRS, IPAG, F-38000 Grenoble, France}

\begin{abstract}
There is mounting evidence that the composition and structure of planetary systems are intimately linked to their birth environments.
During the past decade, several spectral surveys probed the chemistry of the earliest stages of star formation and of late planet-forming disks. However, very little is known about the chemistry of intermediate protostellar stages, i.e. Class I Young Stellar Objects (YSOs), where planet formation may have already begun.
We present here the first results of a 3mm spectral survey performed with the IRAM-30m telescope to investigate the chemistry of a sample of seven Class I YSOs located in the Taurus star-forming region. These sources were selected to embrace the wide diversity identified for low-mass protostellar envelope and disk systems. We present detections and upper limits of thirteen small ($N_{\rm atoms}\leq3$) C, N, O, and S carriers - namely CO, \ce{HCO+}, HCN, HNC, CN, \ce{N2H+}, \ce{C2H}, CS, SO, \ce{HCS+}, \ce{C2S}, \ce{SO2}, OCS - and some of their D, $^{13}$C, $^{15}$N, $^{18}$O, $^{17}$O, and $^{34}$S isotopologues. Together, these species provide constraints on gas-phase C/N/O ratios, D- and $^{15}$N-fractionation, source temperature and UV exposure, as well as the overall S-chemistry. We find substantial evidence of chemical differentiation among our source sample, some of which can be traced back to Class I physical parameters, such as the disk-to-envelope mass ratio (proxy for Class I evolutionary stage), the source luminosity, and the UV-field strength. Overall, these first results allow us to start investigating the astrochemistry of Class I objects, however, interferometric observations are needed to differentiate envelope versus disk chemistry.
\end{abstract}


\section{Introduction} 
\label{sec:intro}

Planets have long been supposed to form by material accretion in mature planet-forming disks, a.k.a. Class II disks, i.e. when disk parental envelopes have been cleared away \citep{lada1984,andrews2005,dullemond2007}.
This evolutionary stage was thus identified as the birthplace of planets, containing both the chemical and mass reservoirs for planet formation. However, during the past decade, evidence started to challenge this assumption. 

First, Class II disk masses are found to be, on average, well below the so-called minimum mass solar nebula \citep[MMSN, ][]{hayashi1981,weidenschilling1977} of 0.01-0.1 M$_{\odot}$ required to form planetary systems such as our own, and more generally below the typical masses of exoplanetary systems \citep[e.g.,][]{barenfeld2016,ansdell2016,law2017,tazzari2017,long2017,eisner2018,cazzoletti2019}. 
Although Class II disk gas and dust masses
could be underestimated due to, for example, poorly-understood CO depletion and/or optical depth effects \citep{yu2017a,yu2017b,zhang2019,andrews2020}, if Class II disk masses are effectively well below the MMSN and, more generally, below the typical masses of exoplanetary systems, planet formation would need either 1) to start at earlier evolutionary stages, i.e. in the Class I stage, where the circumstellar envelope starts to fade away as matter accretes onto the nascent protoplanetary disk and the central protostar \citep[e.g.,][]{harsono2018} ; or 2) to still be fed by a regular replenishment of the protoplanetary disk from the parental molecular cloud \citep[e.g.,][]{kuffmeier2017,manara2018,dullemond2019,kuffmeier2020}. 

Second, recent ALMA observations revealed that most Class II disks and even a few Class I disks \citep[e.g., HL Tau, L1551 NE, and GY 91,][]{yen2017,takakuwa2017,sheehan2018} possess sub-structures such as rings, clumps, spirals and large-scale crescents \citep[e.g.,][]{alma_partnership2015,andrews2018,huang2018,long2018}. 
These sub-structures are somehow puzzling since they are at the same time supposed to {\it(i)} trace planets in the making (i.e., by-products of planet formation) and {\it(ii)} facilitate/initiate planet formation. Even though their primary origin remains to be elucidated, these sub-structures likely indicate that planet formation starts at earlier stages of disk evolution.

Thus, to get the full picture of planet genesis and chemical origins of planet compositions, it is crucial to study earlier evolutionary stages starting with Class I Young Stellar Objects (YSOs), which are still embedded in their parental envelope \citep[e.g.,][]{white2007}, i.e. protostellar envelope and disk systems. In the Class I phase, the disk chemistry is likely influenced both by the infall of fresh envelope material \citep[e.g.,][]{yen2013,sakai2014,jacobsen2019} and by the increasing importance of UV and X-ray emissions from the star \citep[e.g.,][]{feigelson1999,igea1999,bergin2003,jonkheid2004,gorti2009,woitke2010,woitke2016,dutrey2014}.
It is unclear how much of the resulting chemical composition is inherited from their natal molecular cloud versus reset by chemical reprocessing.
To address this question,  we carried out a spectral line survey of a small sample of Class I systems to explore the chemical nature of these objects.

Here, we present IRAM-30m observations of a sample of seven low-mass protostellar systems located in the Taurus star-forming region and selected to encompass a wide range of Class I physical properties. We aim to explore the chemical content of Class I YSOs and the relationship between molecular emission patterns and the physical properties of these objects. 
The source sample and observations are described in Section~\ref{sec:obs}. Our main findings are reported in Section~\ref{sec:results} where, after presenting the different line detections (\S~\ref{subsec:line_det}) and line flux variation across the source sample (\S~\ref{subsec:line_flux_overview}), we focus on line flux ratios proposed to probe C/N/O gas-phase ratios, deuterium fractionation, temperature, UV-field strength, and chemical evolution (\S~\ref{subsec:line_flux_ratios}). In Section~\ref{sec:discussion}, we discuss the chemical differentiation found across our sample, and how it compares to other Class I studies, and to Class 0 and Class II chemistry. We present our conclusions in Section~\ref{sec:conclusion}.

\begin{table*}
\centering
\tiny
\scriptsize
{\caption{Characteristics of the targeted sources sorted by decreasing disk-to-envelope mass ratio}}
{
\begin{tabular}{lccccccccccc}
\hline \hline \noalign {\smallskip}
Source& R.A.$^{(a)}$&Dec.$^{(a)}$& $T_{bol}$& $L{_\star}^{(b)}$ & $ M_{\rm{Env.}}^{(b)}$ &$M_{\rm{Disk}}^{(b)}$ &$M_{\rm{Disk}}^{(b)}$/ & $R_{\rm{Env.}}^{(b)}$ &$R_{\rm{Disk}}^{(b)}$  &V$_{\rm{LSR}}$ & Dist.\\
& (J2000) & (J2000)& (K)&($L_{\odot}$) & ($M_{\odot}$)& (M$_{\odot}$) &$M_{\rm{Env.}}$ &(au)&(au) &(km/s)& (pc)\\
\hline \noalign {\smallskip}
IRAS~04302+2247   & 04:33:16.501 & 22:53:20.400 & 122$^{(c)}$ & 0.4 & 0.017$^{+0.006}_{-0.004}$ & 0.114$^{+0.019}_{-0.026}$
 & 6.7     & 1086   & 244     & 5.5 [1]&161$\pm$3$^{(f)}$\\
IRAS~04295+2251   & 04:32:32.055 & 22:57:26.670 & 270$^{(c)}$ & 0.3 & 0.037$^{+0.008}_{-0.006}$ & 0.018$\pm 0.001$ & 0.49     & 1081   & 127     & 5.3 [1]&161$\pm$3$^{(f)}$\\
IRAS~04365+2535   & 04:39:35.194 & 25:41:44.730 & 164$^{(d)}$ & 2.1 & 0.071$^{+0.035}_{-0.019}$ & 0.030$^{+0.002}_{-0.003}$ & 0.42     & 1829   & 143     & 6.6 [2]&140$\pm$4$^{(f)}$\\
IRAS~04016+2610   & 04:04:43.071 & 26:18:56.390 & 226$^{(d)}$ & 7.0 & 0.023$^{+0.010}_{-0.004}$ & 0.009$\pm 0.001$ & 0.39     & 1446   & 497     & 6.8 [2] &$\sim$140$^{(g)}$\\
IRAS~04166+2706   & 04:19:42.627 & 27:13:38.430 &  75$^{(c)}$ & 0.2 & 0.100$\pm 0.009$ & 0.027$\pm 0.003$ & 0.27     & 1209   & 180     & 6.7 [3]& 160$\pm$3$^{(f)}$\\ 
IRAS~04169+2702   & 04:19:58.449 & 27:09:57.070 & 133$^{(c)}$ & 0.8 & 0.055$^{+0.004}_{-0.005}$ & 0.012$\pm 0.001$ & 0.22     & 672     & 39       & 6.8 [2]& 160$\pm$3$^{(f)}$\\
IRAS~04181+2654A & 04:21:11.469 & 27:01:09.400 &346$^{(e)}$ & 0.3 & 1.234$^{+0.688}_{-0.391}$ & 0.006$\pm 0.001$ & 4.8e-3 & $>20000$ & 47 & 7.1 [1] & 160$\pm$3$^{(f)}$\\
\hline \noalign {\smallskip}
\end{tabular}
    \label{tab:sources}
    \tablenotetext{a}{Right ascension (R.A.) and declination (Dec.) of each source are from the Gaia DR2 catalog \citep{gaia_catalog2018}}
    \tablenotetext{b}{These values correspond to the best-fit parameters derived in \cite{sheehan2017} with their MCMC model, where the uncertainties are the range containing 68\% of the posterior distribution. Systematic errors could be larger.}
    \tablenotetext{c}{From \cite{young2003}}
    \tablenotetext{d}{From \cite{green2013}}
    \tablenotetext{e}{From \cite{beck2007}}
    \tablenotetext{f}{From \cite{galli2019}, using the distance of the molecular clouds where the sources are located (see Section~\ref{subsec:sce_sample}).}
    \tablenotetext{g}{Identified as an outlier in \cite{galli2019}, so we kept here the former distance used for the Taurus star-forming regions.}
    \tablecomments{ [1] Deduced from the rotational transition $N=1-0$ of HCN from this work, [2] Deduced from the rotational transition $N=1-0$ of CN from this work, [3] from \cite{santiago-garcia2009} and in agreement with the one deduced from the rotational transition $N=1-0$ of HCN from this work.}
    }
\end{table*}

\section{Observations}
\label{sec:obs}
\subsection{Source Sample}
\label{subsec:sce_sample}
We selected seven Class I sources with well-characterized physical properties \citep[e.g.,][]{sheehan2017}, located in the nearby star-forming region of Taurus \citep[$d < 200$~pc,][see Table~\ref{tab:sources}]{galli2019}. The targeted sources were chosen to maximize the diversity of physical properties (disk-to-envelope ratio, luminosity and disk size) in order to elucidate their relationship with the disk chemical properties. The coordinates and properties of each source are listed in Table \ref{tab:sources}; the sources span luminosities ranging from 0.3 to 7~$L_\odot$, envelope  masses from 0.02 to 1.2~$M_\odot$ and masses of their embedded nascent disks from 0.01 to 0.1 $M_\odot$. We provide a brief description of our selected sources in the following.

\subsubsection{IRAS~04302+2247}
\label{subsubsec:302}

IRAS~04302+2247 (hereafter I-04302), the so-called Butterfly star \citep{lucas1997}, is a well-known YSO with an edge-on circumstellar disk located near the L1536b dark cloud \citep[e.g.,][]{wolf2003,connelley2010}. Of the sources in our sample, this one has the largest disk-to-envelope mass ratio ($\approx 6.7$), presenting a massive disk, with $M{_{\rm disk}}~\approx~0.114 M_\odot$ and a 240~au radius, and is close to fully dispersing its natal envelope \citep[$M{_{\rm envelope}}~\approx~0.017 M_\odot$,][]{sheehan2017}. This indicates the advanced evolutionary stage of this YSO with respect to the six other sources of our sample (see Table~\ref{tab:sources}).

\subsubsection{IRAS~04295+2251}
\label{subsubsec:295}
IRAS~04295+2251 (hereafter I-04295), also known as L1536~IRS, is classified as a Class I object according to its bolometric temperature of $T_{bol}=270$~K \citep{young2003}. Although \cite{chiang1999} suggested that it could be an edge-on disk, later on observations rather suggest an inclined disk \citep{furlan2008,eisner2012,sheehan2017}. Although its disk-to-envelope mass ratio is $\approx0.5$, one order of magnitude lower than for I-04302, it is, however, the second largest of our sample. Thus, contrary to I-04302, this indicates that the disk is still well embedded in its natal envelope, with $M{_{\rm disk}} \approx 0.02 M_\odot$ and a 130~au radius.

\subsubsection{IRAS~04365+2535}
\label{subsubsec:365}
IRAS~04365+2535 (hereafter I-04365), also known as TMC1A, is located at the center of the L1534 dark cloud \citep{benson1989}. I-04365 shows an inclined Keplerian rotating disk \citep[e.g.,][]{harsono2014,aso2015,sheehan2017}, with a disk-to-envelope mass ratio of $\approx0.4$, third largest of our sample, with $M{_{\rm disk}} \approx 0.04 M_\odot$ and a 140~au disk radius.
The disk is driving a wind \citep{bjerkeli2016} and large-scale outflow \citep{chandler1995,aso2015} probably both at the origin of a forming cavity in the surrounding envelope.

\subsubsection{IRAS~04016+2610}
\label{subsubsec:016}
IRAS~04016+2610  (hereafter I-04016), also known as L1489~IRS, is a Class I system \citep[$T_{bol}= 226$~K,][]{green2013} 
likely located behind a foreground starless cloud \citep[e.g.,][]{hogerheijde2000,brinch2007a}. This source shows evidences of a disk in Keplerian rotation misaligned from its surrounding envelope \citep{hogerheijde2001,brinch2007b} and a bipolar outflow 
\citep{myers1988,hogerheijde1998,yen2014}. Although this source has the largest disk of our sample, it also has
one of the smallest disk masses ($M{_{\rm disk}}=0.009 M_\odot$, see Table~\ref{tab:sources}), with a disk-to-envelope mass ratio $\approx 0.4$ indicative of the embedded nature of this source.

\subsubsection{IRAS~04166+2706}
\label{subsubsec:166}
IRAS~04166+2706 (hereafter I-04166) is a YSO located in the molecular filament of the L1495/B213 Taurus \citep{hacar2013}, harbouring highly collimated jet and bipolar outflow \citep{bontemps1996,tafalla2004,santiago-garcia2009}. Claimed to be a Class 0 system by some \citep[e.g.,][]{tafalla2004,tafalla2017,wang2014,wang2019} and a Class I source by others \citep[e.g.,][]{young2003,furlan2008,eisner2012,sheehan2017}, it is probably in transition between these two stages. This is in agreement with the fact that I-04166 is the source with the lowest luminosity ($\approx 0.2 L_{\odot}$) and bolometric temperature ($T_{bol}=75$~K) in our sample (see Table~\ref{tab:sources}).

\subsubsection{IRAS~04169+2702}
\label{subsubsec:169}
IRAS~04169+2702 (hereafter I-04169), located in the molecular filament of the L1495/B213 Taurus \citep{hacar2013}, is a Class I YSO \citep[$T_{bol}=$133~K,][]{young2003}
also characterized by a bipolar outflow \citep{bontemps1996} found to be perpendicular to an elongated flattened envelope of about 2200~au $\times$
1100~au in size area \citep{ohashi1997}. A compact $\sim 40$~au radius disk is embedded in this huge envelope \citep{sheehan2017}, with a disk-to-envelope mass ratio $\approx 0.2$.

\subsubsection{IRAS~04181+2654A}
\label{subsubsec:181}
IRAS~04181+2654A (hereafter I-04181), located in the molecular filament of the L1495/B213  Taurus \citep{bontemps1996,davis2010}, is the least characterized source of our sample due to its lack of submillimeter study. It is part of a wide binary system, with a deeper embedded B component lying 31'' away \citep[i.e., $\sim 4300$ au away at the distance of Taurus,][]{furlan2008}. The A component is closer to a dense core and shows two bipolar outflows \citep{davis2010}.

\subsection{Description of the observations}
\label{subsec:desc_obs}

The observations were performed at the IRAM-30m telescope in a single observing run of 49.5h from June 19th to June 23rd, 2019. Each of the seven sources targeted for this study was observed with the spectral line Eight MIxer Receivers (EMIR) in band E090 (3mm), with an almost complete coverage of the 3mm band, covering $72.0 -79.9$~GHz and $84.2-115.5$~GHz, using the wide mode of the Fast Fourier Transform Spectrometers (FTS) \citep{klein2012} at a spectral resolution of 200~kHz (i.e., $\sim 0.5-0.8$ km.s$^{-1}$), with an average sideband rejection of -16dB \citep{carter2012}. The observations were carried out in Wobbler Switching (WSW) mode, with a throw of 180'', to ensure a flat baseline across the spectral bandwidth observed. Frequency Switching (FSW) mode was also used to check if the absorption features appearing on the strongest molecular spectra (mainly $^{12}$CO and \ce{HCO+}) were due to plausible contamination from the OFF positions or probing different physical components of the targeted sources. More observational time would have been required to also check the absorption features appearing on fainter lines with this method, but these first FSW checks provide some clues on the plausible origins of the absorption features observed, as described in Section~\ref{sec:results}. Moreover, the wobbler OFF position is given as a single fixed relative offset in azimuth with respect to the target position. This results in a slightly different absolute position in the sky of the OFF position as the Earth rotates. Thus, contamination lines can be checked in WSW data by analyzing if such lines are moving or disappearing during the observation scans. We used this technique to explore the origins of a handful of lines observed in absorption instead of emission (see Section~\ref{subsec:line_det}).

Primary pointing and focus were done on Uranus and secondary focus were done on quasar 0439+360 and on Venus. Pointing was checked every 1.5 hours on these same two nearby sources. 
The average pointing corrections were between 3" and 4", i.e. 5 to 10 times smaller than the IRAM beam at the observed frequencies, which varies from 34'' at 72~GHz to 22'' at 110~GHz. The averaged achieved rms is $\approx 5$~mK.

The observational data were reduced and analysed using the \texttt{CLASS/GILDAS} software package \citep{pety2005,gildasteam2013} and the line identification and fitting was performed with the \texttt{CASSIS}\footnote{http://cassis.irap.omp.eu} software.

\section{Results}
\label{sec:results}
\begin{figure*}
    \centering
    \includegraphics[scale=0.7]{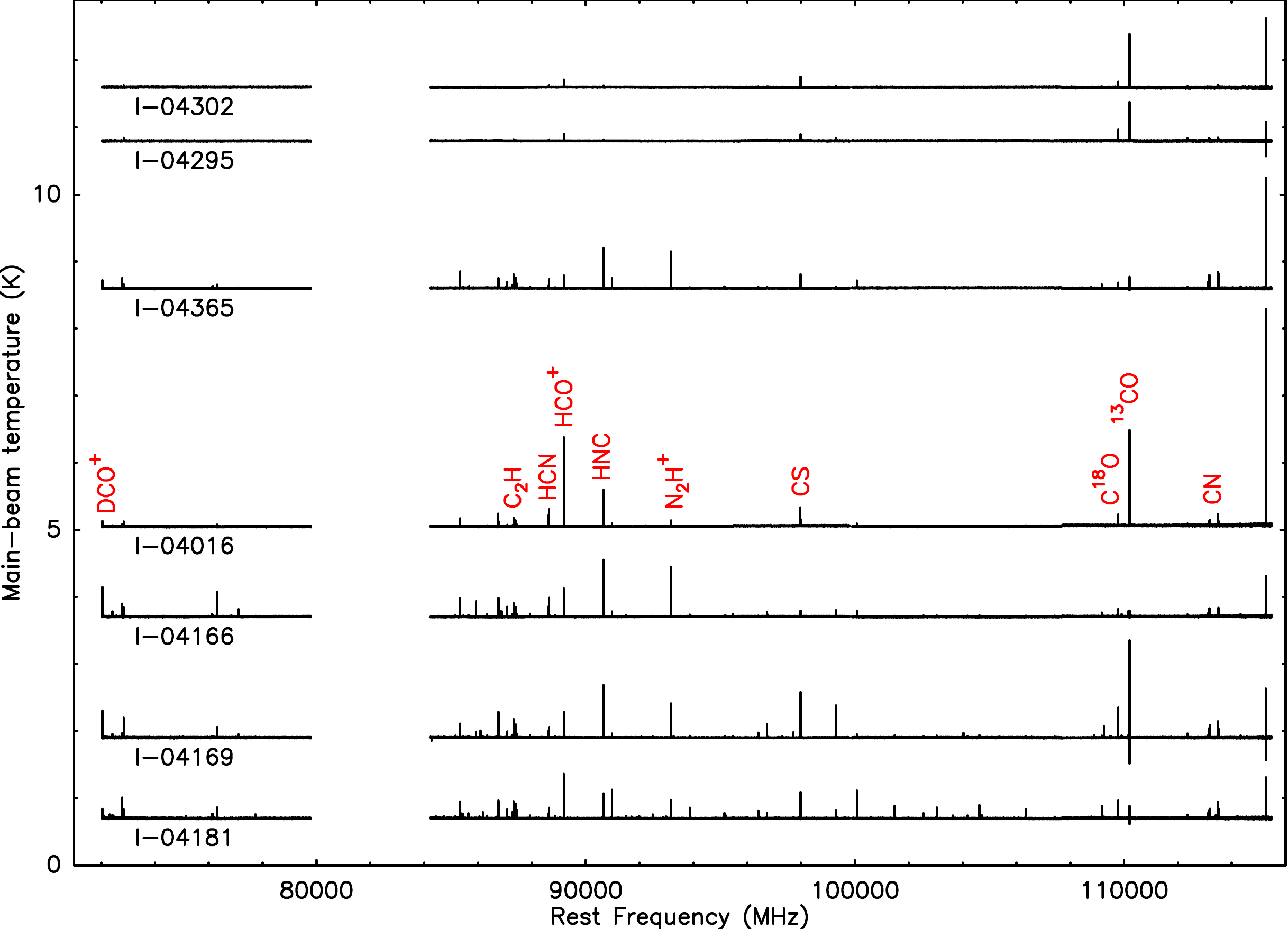}
    \caption{Full 3mm spectral line survey obtained with the IRAM-30m (project ID:014-19), toward each of the seven Class I objects targeted in this work, and sorted by decreasing disk-to-envelope mass ratios (from top to bottom). The gap from $\sim 80$ to $\sim 84$ GHz was not covered by our observations. Ten of the major species detected in this work are presented in red.}
    \label{fig:all_sces_full_3mm}
\end{figure*}

In this section, we present the line detections of small (N$_{\rm{atoms}}\leq3$) molecules in our 3mm spectral survey of seven Class I YSOs and explore their relationship with one another and with source properties across the sample  -- detections of larger molecules will be presented in a future article.

\subsection{Line Detections}
\label{subsec:line_det}

\begin{figure*}
    \centering
    \includegraphics[scale=1.2]{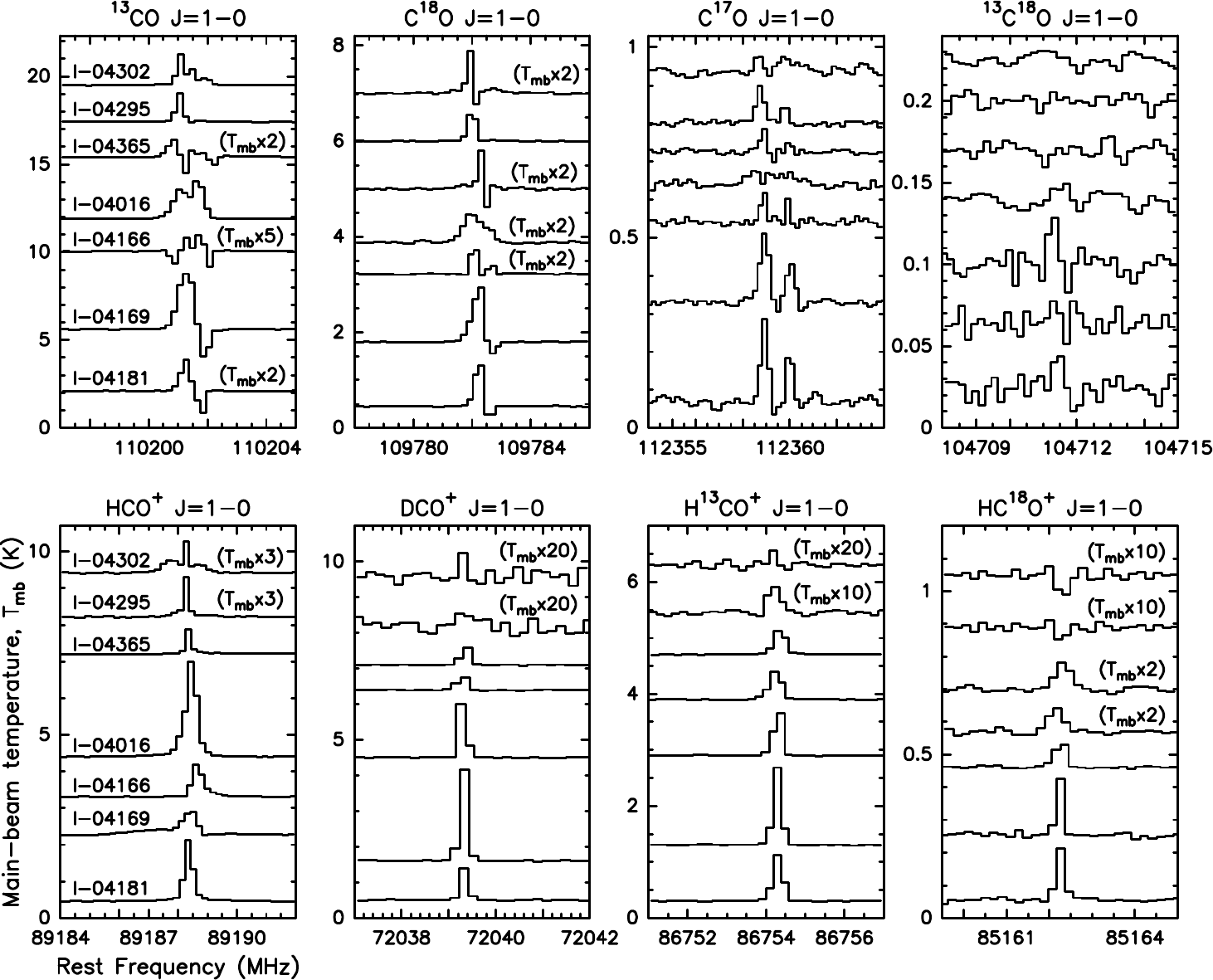}
    \caption{Spectra of the transition $J=1-0$ of
    $^{13}$CO, C$^{18}$O, C$^{17}$O, $^{13}$C$^{18}$O ({\it upper panels}), \ce{HCO+}, \ce{DCO+}, \ce{H^{13}CO+}, and \ce{HC^{18}O+} ({\it lower panels}) observed toward each of the seven Class I sources targeted in this work. Spectra are sorted by decreasing disk-to-envelope mass ratios from top to bottom in each panel, and by decreasing intensity in each row of panels.}
    \label{fig:30m_CO_HCOp_isotopologues}
\end{figure*}

\begin{figure*}[!h]
    \centering
    \includegraphics[scale=1.2]{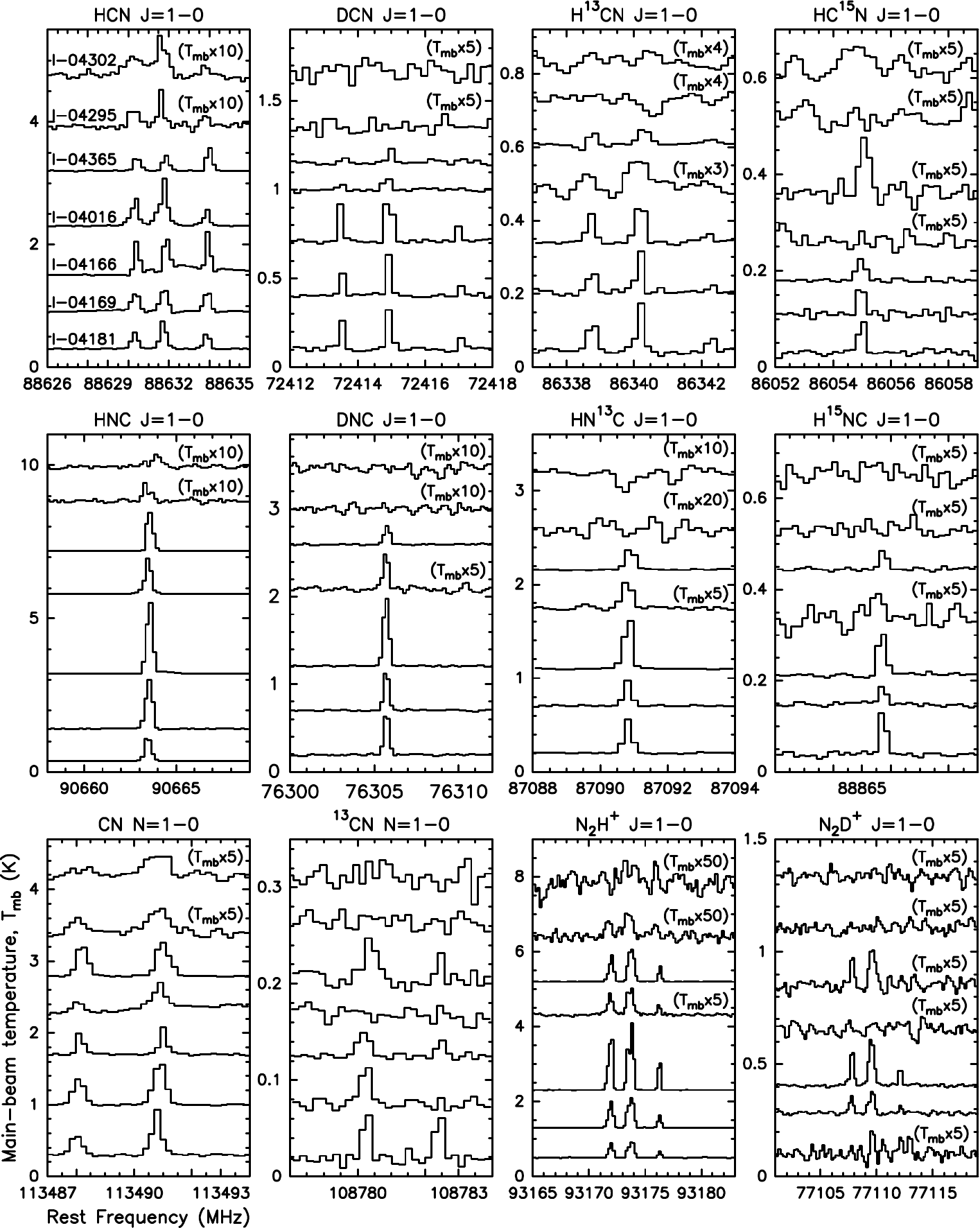}
    \caption{Spectra of the transition $1-0$ of HCN, DCN, \ce{H^{13}CN}, \ce{HC^{15}N} ({\it upper panels}), HNC, DNC, \ce{HN^{13}C}, \ce{H^{15}NC} ({\it middle panels}), \ce{N2H+}, \ce{N2D+}, CN and $^{13}$CN ({\it lower panels}) observed toward each of the seven Class I sources of our sample. Spectra are sorted by decreasing disk-to-envelope mass ratios from top to bottom in each panel, and by decreasing intensity in each row of panels.}
    \label{fig:30m_nitriles}
\end{figure*}

\begin{figure}
    \centering
    \includegraphics[scale=1.2]{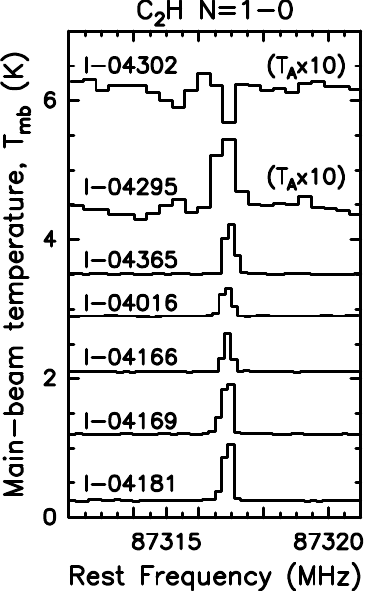}
    \caption{Spectra of the strongest component of \ce{C2H} $N=1-0$ observed toward each of the seven targeted sources of our sample, sorted by decreasing disk-to-envelope mass ratios from top to bottom in each panel.} 
    \label{fig:30m_c2h}
\end{figure}

\begin{figure*}
    \includegraphics[scale=1.2]{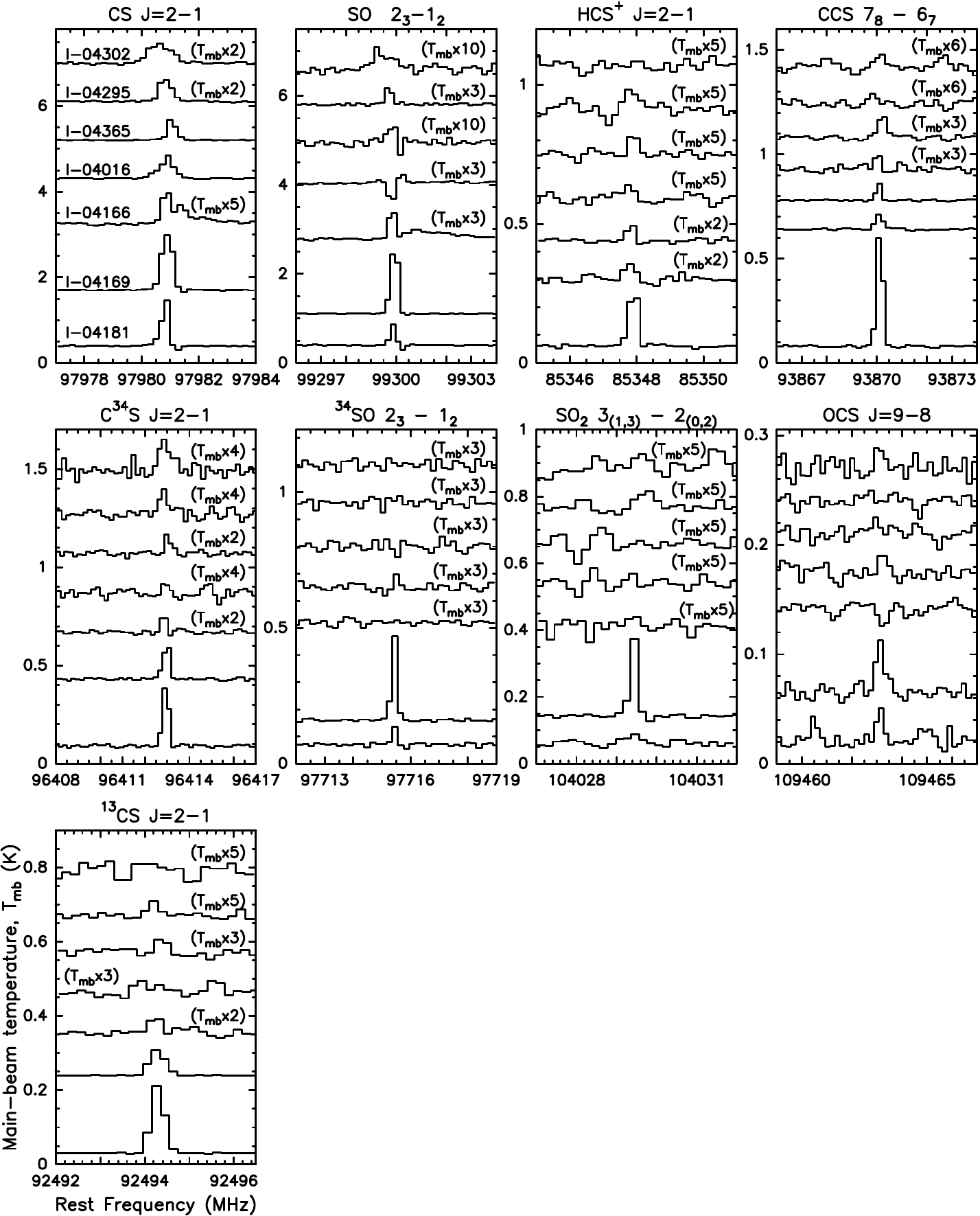}
    \caption{Spectra of the strongest component of CS, SO, \ce{HCS+}, CCS, \ce{^{13}CS}, \ce{C^{34}S}, \ce{^{34}SO}, \ce{SO2}, and OCS observed toward each of the seven Class I sources of our sample. Spectra are sorted by decreasing disk-to-envelope mass ratios from top to bottom in each panel.}
    \label{fig:30m_S-species}
\end{figure*}

Figure~\ref{fig:all_sces_full_3mm} presents the full 3mm spectra obtained toward each of the seven Class I sources targeted in this work, sorted by decreasing disk-to-envelope mass ratios. All seven sources present a large number of spectral lines, but the line intensity and richness decreases with disk-to-envelope mass ratio.
We detected thirty molecules with no more than 3 atoms toward three or more sources in our sample, where our detection criterion is set to be $\geq 5\sigma$ for the integrated intensity. Table~\ref{tab:spectro} lists the most intense line of each species, and the same lines are presented in Figures~\ref{fig:30m_CO_HCOp_isotopologues} to ~\ref{fig:30m_S-species}.

Figure~\ref{fig:30m_CO_HCOp_isotopologues} displays spectra for the transitions $J=1-0$ of all CO and \ce{HCO+} isotopologues observed, with the exception of $^{12}$CO, which will be discussed in a forthcoming study. $^{13}$CO, C$^{18}$O and C$^{17}$O are detected toward all Class I sources, while $^{13}$C$^{18}$O is only seen in the four sources with the smallest disk-to-envelope mass ratios. Since absorption features are seen in $^{13}$CO and C$^{18}$O, which could be indicative of multiple velocity components along the line-of-sight or simply due to the optical thickness of the lines, in the remainder of this work we use C$^{17}$O as a proxy for CO. HCO$^+$, DCO$^+$, and H$^{13}$CO$^+$ are detected toward all sources, but the fainter isotopologue HC$^{18}$O$^+$ is only detected in emission in 5/7 sources. Faint absorption features are detected in the remaining two disk-dominated sources but are likely due to contamination from the OFF positions, because these features only appear in a few scans of the WSW observations.

Nitrogen-bearing species are presented Fig.~\ref{fig:30m_nitriles}, with four isotopologues of HCN and HNC each (i.e., main, D, $^{13}$C and $^{15}$N  isotopologues), and two of CN and \ce{N2H+} each, (i.e., main and $^{13}$C, and main and D isotopologues, respectively). The main isotopologues of these four molecules are detected in all seven sources, while the rarer isotopologues are mainly seen in the low disk-to-envelope mass ratio sources. In addition, HN$^{13}$C is seen in absorption toward I-04302, but we suspect this to be a foreground cloud contamination. We should also note that multiple velocity components are seen in \ce{N2D+} toward I-04181 and in \ce{N2H+} toward I-04302. These data will need higher resolution observations to be further explored. In the remainder of the present study, we focus on the narrow, central emission or absorption feature toward each source from which first general overview results of the chemistry of Class I objects can be straightforwardly derived.

Figure~\ref{fig:30m_c2h} presents the spectra for the sole 3-atom hydrocarbon observed, \ce{C2H} $J=1-0$. It is detected toward all sources, but is observed in absorption toward I-04302. Why some lines are seen in absorption toward this source is unclear. Contrary to the HC$^{18}$O$^+$ $1-0$ line, checking the WSW observation scans one by one suggests that it is less likely that these absorption features are due to cloud contamination. We speculate that the line could originate from absorption in a cold envelope against optically thick dust in the edge-on disk of this system, but further observations are required to check their origin.

Figure~\ref{fig:30m_S-species} presents nine S-bearing species of interest, three isotopologues of CS (2-1) (including the main, $^{34}$S and $^{13}$C isotopologues), two isotopologues of SO ($2_3$-$1_2$) (including the main and $^{34}$S isotopologues), HCS$^+$ (2-1), CCS ($7_8$-$6_7$), OCS (9-8), and \ce{SO2}($3_{1,3}$-$2_{0,2}$).  Absorption, self-absorption, or multiple velocity components are apparent for CS toward I-04166, and in both SO isotopologues toward I-04181, I-04016 and I-04365. In these cases, we integrated the emission of the central, narrow component, as indicated above (see Table~\ref{tab:intensities}).

\begin{table}[]
    \centering
    \footnotesize
    \caption{Properties of all detected lines from CDMS$^{(a)}$}
    \begin{tabular}{lcccccc}
    \hline \hline
         Molecule &Frequency  &$E_{up}$ & $g_{up}$ &$A_{ij}$ & Transition\\
         &(MHz)& (K) & & (s$^{-1}$) \\
    \hline
      $^{13}$CO & 110201.354 & 5.3 & 6 & 6.33e-08 & 1 -- 0 \\
      C$^{18}$O&  109782.173 & 5.3 & 3 & 6.27e-08  & 1 -- 0  \\
      C$^{17}$O$^{(\ddagger)}$ & 112358.982  & 5.4 &   8  &6.70e-08    &  1 4 -- 0 3\\
      $^{13}$C$^{18}$O & 104711.404 & 5.0 & 4 & 5.46e-08 &  1 2 -- 0 1  \\
      \hline
      HCO$^+$ &  89188.525 &  4.3   & 3  &4.19e-05  &       1 -- 0 \\
      DCO$^+$ & 72039.312  &     3.5  &  3 & 2.21e-05  &    1 -- 0 \\
      H$^{13}$CO$^+$ & 86754.288  &  4.2   & 3 & 3.85e-05 & 1 -- 0 \\
      HC$^{18}$O$^+$ & 85162.223  &  4.1   & 3 & 3.65e-05 & 1 -- 0 \\
      \hline
     HCN$^{(\ddagger)}$
         &  88631.847  & 4.3  &  5  &2.43e-05& 1 2 -- 0 1  \\ 

     DCN$^{(\ddagger)}$ &  72414.905  & 3.5   & 5  & 1.31e-05 &  1 2 -- 0 1  \\
     H$^{13}$CN$^{(\ddagger)}$ & 86339.921 & 4.1 & 9 & 2.23e-05              & 1 -- 0 \\
     HC$^{15}$N & 86054.966 & 4.1 & 3 & 2.20e-05  & 1 -- 0\\
      \hline
     HNC & 90663.568 &4.4 &3 &2.69e-05 & 1 -- 0\\
     DNC &76305.700 &3.7  &3  &1.60e-05 & 1 -- 0  \\
     HN$^{13}$C &  87090.825 & 4.2 & 3 & 2.38e-05  & 1 -- 0 \\
     H$^{15}$NC & 88865.715 & 4.3 & 3 & 1.98e-05  & 1 -- 0  \\
 \hline
CN$^{(\ddagger)}$  &   113490.970  &     5.4   & 6  & 1.19e-05     &  1 0 2 3 -- 0 0 1 2\\
      $^{13}$CN$^{(\ddagger)}$ &108780.201 & 5.2  &  7 & 1.05e-05   &     1 2 2 3 -- 0 1 1 2  \\
    \ce{N2H+}$^{(\ddagger)}$ 
    &  93173.700   &  4.5  & 15 & 3.63e-05   & 1 2 -- 0 1\\
    \ce{N2D+}$^{(\ddagger)}$ 
    & 77109.610  &  3.7  &  5  &2.06e-05   &  1 2 -- 0 1   \\
    \hline
    \ce{C2H}$^{(\ddagger)}$ &87316.898  &4.2   & 5 & 1.53e-06     &     1 2 2 -- 0 1 1   \\
    \hline
          CS   &  97980.953 &  7.1  &  5  &1.68e-05  &2 0 -- 1 0  \\
      $^{13}$CS & 92494.308 & 6.7 & 10 & 1.41e-05 &  2 0 -- 1 0 \\ 
      C$^{34}$S & 96412.950 & 6.9 & 5  & 1.60e-05 &  2 0 -- 1 0  \\
      SO  
          & 99299.870 & 9.2. & 7 & 1.13e-05 & 2 3 -- 1 2 \\
    $^{34}$SO     &  97715.317   &  9.1  &  7 & 1.07e-05   &         2 3 -- 1 2  \\
      \ce{HCS+} &85347.890 &  6.1  &  5 & 1.11e-05   &  2 -- 1  \\  
      CCS     
              & 93870.107 & 19.9 &  17 & 3.74e-05 &  7 8 -- 6 7\\
     \ce{SO2} 
      &104029.418  &7.7 & 7 & 1.01e-05 &         3 1 3 -- 2 0 2   \\  
      OCS  & 109463.063 &  26.3  & 19  & 3.70e-06   &           9 -- 8  \\ 
      \hline
      \end{tabular}
      \tablenotetext{a}{https://cdms.astro.uni-koeln.de/cdms/portal/; \citep{cdms2001,cdms2005}}
      \tablenotetext{\ddagger}{For species with hyperfine structure, properties are only given for the main component.}
    \label{tab:spectro}
\end{table}

\subsection{Integrated line Fluxes Across Our Source Sample}
\label{subsec:line_flux_overview}
\begin{deluxetable*}{|l|ccccccc|}
\tablecaption{Integrated line intensities in mK km/s of the twenty-seven species of interest for this work.\\
\label{tab:intensities}}
\tablewidth{1200pt}
\tabletypesize{\footnotesize}
\tablehead{\diagbox[innerwidth=2cm,innerrightsep=-3pt]{Molecule}{Source} & \colhead{I-04302} & \colhead{I-04295} & \colhead{I-04365}& \colhead{I-04016}& \colhead{I-04166}& \colhead{I-04169}& \colhead{I-04181}} 
\startdata
   C$^{17}$O$^{(\ddagger)}$    &$26.7\pm 2.3$    &$33.0\pm 1.6$    &$21.9\pm 1.6$      &$31.6\pm 3.9$    &$21.7\pm 1.6$    &$76.6\pm 1.9$    &$100.5\pm 4.2$    \\
    HCO$^+$   &$177.8\pm 2.5$    &$91.7\pm 2.8$   &$165.6\pm 3.1$    &$1228.1\pm 7.3$   &$398.5\pm 5.2$  &$346.6\pm 15.4$    &$632.6\pm 5.9$                     \\
    DCO$^+$   &\it{6.4 $\pm$ 1.8}~$^{(a)}$   &$<10.1^{(b)}$ 
    &$136.9\pm 2.7$     &$109.8\pm 2.9$   &$363.0\pm 3.3$   &$576.4\pm 4.7$    &$199.2\pm 5.3$     \\
 H$^{13}$CO$^+$     &\it{3.7 $\pm$ 1.0}    &$25.5\pm 1.0$   &$159.0\pm 1.0$     &$200.2\pm 1.1$   &$238.6\pm 1.1$   &$387.8\pm 1.3$    &$271.4\pm 1.3$                     \\
 HC$^{18}$O$^+$     & ...$^{(c)}$ & ...  &$20.1\pm 1.4$      &$17.9\pm 1.3$    &$28.7\pm 1.1$    &$40.5\pm 1.3$     &$42.3\pm 1.3$                     \\
    HCN$^{(\ddagger)}$    &$77.7\pm 3.1$    &$38.4\pm 2.5$   &$311.7\pm 1.9$     &$603.1\pm 5.2$   &$492.7\pm 3.9$   &$460.5\pm 6.0$    &$364.8\pm 2.5$  \\
    DCN$^{(\ddagger)}$    &$<11.5$    &$<11.5$    &$30.1\pm 4.0$       &$29.0\pm 3.4$   &$147.5\pm 2.8$    &$87.6\pm 4.3$     &$63.8\pm 6.6$   \\
    H$^{13}$CN$^{(\ddagger)}$    &$<10.5$    &$<10.5$    &$33.3\pm 3.5$       &$27.2\pm 3.3$    &$62.3\pm 1.9$    &$49.3\pm 2.5$     &$71.0\pm 3.3$   \\
    HC$^{15}$N    &$<10.6$     &$<4.7$     &$7.0\pm 1.3$        &\it{8.3 $\pm$ 1.8}    &$11.2\pm 1.6$     &\it{9.9 $\pm$ 2.9}     &$20.0\pm 1.4$   \\
   HNC  &$23.9\pm2.7$ & $20.8 \pm 2.2$   &$558.8 \pm 0.05$  &$497.9 \pm 1.8$ &$969.1 \pm 6.5$   &$742.8 \pm 2.1$    &$348.9 \pm 1.5$   \\
   DNC     &$<4.6$     &$<4.6$    &$93.2\pm 1.9$       &$34.4\pm 1.9$   &$342.4\pm 1.8$   &$180.3\pm 2.7$    &$198.6\pm 2.5$   \\
   HN$^{13}$C  & ... & ...   &$82.1\pm 1.6$       &$19.1\pm 1.1$   &$173.4\pm 1.4$    &$75.8\pm 1.8$    &$115.7\pm 1.3$   \\
   H$^{15}$NC     &$<4.7$     &$<4.7$    &$13.5\pm 1.1$      &\it{5.5 $\pm$ 1.9} &$34.4\pm 1.1$    &$12.3\pm 1.0$     &$29.0\pm 1.2$   \\
    CN  &$279.5\pm 22.5$  &$259.2\pm 15.2$  &$1682.0\pm 9.3$     &$698.7\pm 15.8$   &$780.0\pm 8.9$  &$1354.2\pm 7.7$  &$1178.7\pm 10.0$   \\
      $^{13}$CN     &$<7.2$     &$<4.8$    &$30.6\pm 4.1$       &\it{11.8 $\pm$ 2.8} &{\it 6.6 $\pm$ 1.7}    &$16.1\pm 1.8$     &\it{21.9 $\pm$ 6.4}   \\
     N$_2$H$^+$     &$<5.9$    &$17.0\pm 1.8$  &$1044.0\pm 0.1$      &$149.3\pm 3.3$  &$1906.4\pm 0.2$   &$997.2\pm 1.1$    &$520.8\pm 4.5$   \\
     N$_2$D$^+$     &$<4.6$     &$<8.0$    &$40.2\pm 3.3$       &$<13.8$   &$202.5\pm 2.7$   &$105.2\pm 4.5$     &$>15.9\pm 3.1^{(d)}$   \\
         C$_2$H    & ...    &$53.3\pm 4.3$   &$922.2\pm 2.7$      &$460.5\pm 2.7$   &$589.8\pm 3.7$   &$847.8\pm 4.6$   &$1193.8\pm 5.9$   \\
    CS   &$228.5\pm 4.5$   &$143.2\pm 2.7$   &$180.7\pm 2.3$      &$312.0\pm 3.5$    &$>98.6\pm 4.1$   &$660.5\pm 1.3$    &$335.6\pm 4.7$   \\
    C$^{34}$S    &\it{15.5 $\pm$ 3.3}    &\it{10.8 $\pm$ 2.3}    &$14.5\pm 1.6$        &\it{8.9 $\pm$ 2.2}    &$15.2\pm 1.6$    &$67.2\pm 2.0$    &$104.9\pm 2.2$   \\
    $^{13}$CS     &$<2.9$     &\it{7.4 $\pm$ 1.6} &\it{4.4 $\pm$ 1.1}        & $<3.8$  &\it{7.4 $\pm$ 1.7} &$20.5\pm 1.4$     &$65.5\pm 1.7$   \\
             SO    &$37.0\pm 4.2$    &$44.3\pm 2.4$    &$>16.2\pm 2.0$       &$>23.1\pm 5.3$    &$68.1\pm 6.4$   &$564.0\pm 1.8$    &$>129.0\pm 5.1$   \\
      $^{34}$SO &$<3.0$ &$<3.0$ &$<3.0$ &\it{4.4 $\pm$ 1.3} &$<3.0$    &$78.6\pm 1.3$     &$14.5\pm 1.6$   \\
        HCS$^+$ &$<2.9$ &\it{5.9 $\pm$ 1.4} &\it{4.6 $\pm$ 1.2} &\it{7.4 $\pm$ 2.0} &$9.3\pm 1.1$    &$10.9\pm 1.9$     &$69.4\pm 1.8$   \\
         C$_2$S     & $<3.8$    &\it{3.3 $\pm$ 1.0}& $15.3\pm 1.3$        &$7.8\pm 1.3$    &$22.9\pm 0.8$    &$28.0\pm 1.2$    &$181.6\pm 1.6$   \\
         SO$_2$     &$<4.2 $ &\it{8.8 $\pm$ 1.8} &$<2.4$        &$<2.4$     &$<2.4$    &$54.4\pm 1.9$      &\it{8.7 $\pm$ 2.8}   \\
            OCS     &$<4.3$ &$<3.0$     &$<3.0$        &$9.8\pm 2.2$     &...    &$23.3\pm 2.8$&\it{9.5 $\pm$ 2.5}          \\
\enddata
\tablenotetext{a}{Tentative detection are indicated in italics, where our tentative detection criterion is set to be $5 \sigma > S/N \geq 3\sigma$.}
\tablenotetext{b}{Upper limits are given for the non-detection cases.}
\tablenotetext{c}{ "..." indicates the lines seen in absorption that we suspect to be contamination.}
\tablenotetext{d}{In case of significant absorption features within emission ones, only lower limits are given.}
\tablenotetext{\ddagger}{For species with hyperfine structure, when several hyperfine components are detected we integrated the intensity for all of the components.}
 \tablenotetext{\ddagger\ddagger}{~$^{13}$CN, is the only species with an hyperfine structure for which the integrated intensity is derived only from the main component. Given how faint the intensities of the other component are compared to the main one, this is a reasonable assumption that does not affect the results presented in this work.}
\end{deluxetable*}

\begin{figure*}
    \centering
    \includegraphics[scale=0.44]{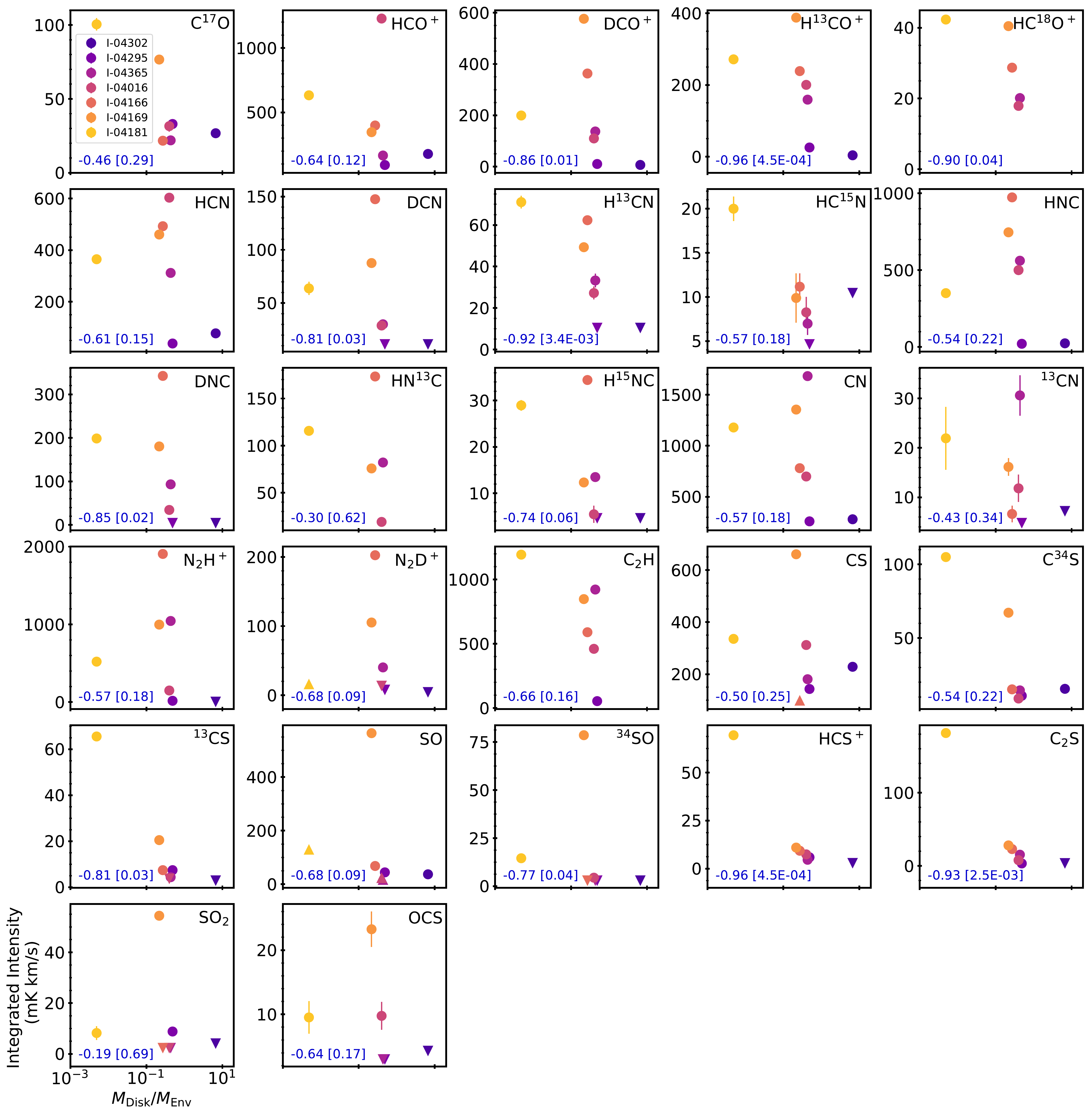}
    \caption{Integrated line intensities of the species of interest presented in this study and represented as function of the disk-to-envelope mass ratio across our sample of Class I YSOs. Error bars are represented by the vertical line segments. Upper limits are represented by the downward triangles and lower limits by the upward triangles. Spearman correlation coefficients are indicated in the lower left corner of each panel, with corresponding p-value in brackets. These statistical parameters are described in \S~\ref{subsec:line_flux_overview}. Note that correlation calculations include upper limits.}
    \label{fig:all_species_flux}
\end{figure*}

All line profiles were fitted by a single Gaussian to extract the integrated line intensities. A line is considered detected when $S/N \geq 5\sigma$, and tentatively detected when $5 \sigma > S/N \geq 3\sigma$. The upper limits on the integrated line intensities were derived assuming a Gaussian shape and a line width of $\sim0.5$ km.s$^{-1}$, similar to the typical line width detected for the other molecules in these sources.
Table~\ref{tab:intensities} presents the integrated line intensities derived for all species presented in this work, except for the CO isotopologues, where we only focused on C$^{17}$O as mentioned earlier.~Figure~\ref{fig:all_species_flux} presents the integrated line intensities of all twenty-seven species {\it versus} disk-to-envelope mass ratio. For most  species, line intensities vary by 1-2 orders of magnitude across the source sample, even within the central narrow range of disk-to-envelope mass ratios varying from $\sim0.2$ to $\sim0.5$, i.e. excluding I-04181 and I-04302 (see Table~\ref{tab:sources}).

A majority of lines are the weakest in the disk-dominated sources, but otherwise there is a range of trends. 33\% (9/27) of the targeted lines are the brightest toward the source with smallest disk-to-envelope mass ratio I-04181, namely C$^{17}$O, HC$^{18}$O$^+$, H$^{13}$CN, HC$^{15}$N, C$_2$H, C$^{34}$S, HCS$^+$, C$_2$S, $^{13}$CS. Most of these lines are typically associated with dense gas \citep[e.g.,][]{bergin1997,vandishoeck1998,tafalla2002,magalhaes2018,delavillarmois2019}, and their overabundance in I-04181 may simply reflect more material in this envelope-dominated line-of-sight.

26\% (7/27) of the lines are instead brightest toward the source with the second smallest disk-to-envelope mass ratio, I-04169, namely DCO$^+$, H$^{13}$CO$^+$, CS, SO, $^{34}$SO, SO$_2$, OCS. Note that this group contains all of the oxygen containing S-species, which is probably suggesting that shocked regions play a non-negligible role. An equal number of lines are brightest toward the source with the third smallest disk-to-envelope mass ratio, I-04166, namely HNC, DCN, DNC, HN$^{13}$C, H$^{15}$NC, N$_2$H$^+$, N$_2$D$^+$. These are all species associated with cold gas, which is consistent with the relatively low luminosity and low bolometric temperature of this source (see Table~\ref{tab:sources}), likely resulting in a large cold gas reservoir in the surrounding disk and envelope. Note that the only deuterated species that is not the brightest in this source is DCO$^+$, which may reflect different excitation conditions or chemistry for DCO$^+$ compared to the other deuterated species, such as its relationship with the depletion of CO \citep{caselli2002,oberg2015}.

HCN and \ce{HCO+} lines are the brightest toward the most luminous source, I-04016, with increased width likely reflecting their optically thickness, as confirmed by our line optical depth study presented in Appendix~\ref{app:opacities}. 
Finally, CN and $^{13}$CN are the brightest toward I-04365, perhaps indicating that this lower disk-to-envelope mass ratio source is more influenced by stellar UV illumination, since CN can be enhanced through photodissociation of HCN driven by Ly$\alpha$ radiation from the central star \citep[e.g.,][]{bergin2003}. Note that none of the targeted lines of this work are found to be the brightest in the two sources with the highest disk-to-envelope ratios, I-04295 and I-04302.  

For each line presented here, we calculated the Spearman's rank correlation coefficient, $\rho_s$, between the integrated line intensity and the disk-to-envelope mass ratio across our Class I source sample, as shown in Fig.~\ref{fig:all_species_flux}. This correlation method measures monotonic relationships, which is a reasonable approach for our case study where there is no reason to expect strictly linear trends. Moreover, this method is less sensitive to outliers than other methods such as the Pearson correlation method. Nine species display statistically significant anti-correlations with disk-to-envelope mass ratio, i.e. with $|\rho_s|$ $\gtrsim 0.8$ and p-values $\leq 0.05$ \citep[where the p-values are used as proxies to assess the statistical significances, e.g.,][]{wasserstein2016}:
HCS$^+$, 
H$^{13}$CO$^+$, 
C$_2$S, 
H$^{13}$CN, 
HC$^{18}$O$^+$, 
DCO$^+$, 
DNC, 
$^{13}$CS, 
and DCN. 

\subsection{Integrated Line Flux Ratios}
\label{subsec:line_flux_ratios}

In the following sub-section, we explore the relationship between subsets of species that have been proposed to trace C/N/O gas-phase ratios, deuterium fractionation, temperature, UV-field strength, and evolutionary stage. We note that in the optically thin limit of local thermodynamic equilibrium (LTE), the integrated line flux ratio of two co-spatial molecules presenting similar upper state energies is roughly proportional to their column density ratio. This assumption is, however, only reasonable for a subset of the observed species, since many of the brightest lines are optically thick (see Appendix~\ref{app:opacities}).

\subsubsection{CO, C$_2$H and HCN: Probing C/N/O Elemental Ratios?}
\label{subsec:CNO_ratios}

\begin{figure*}
    \centering
    \includegraphics[scale=0.62]{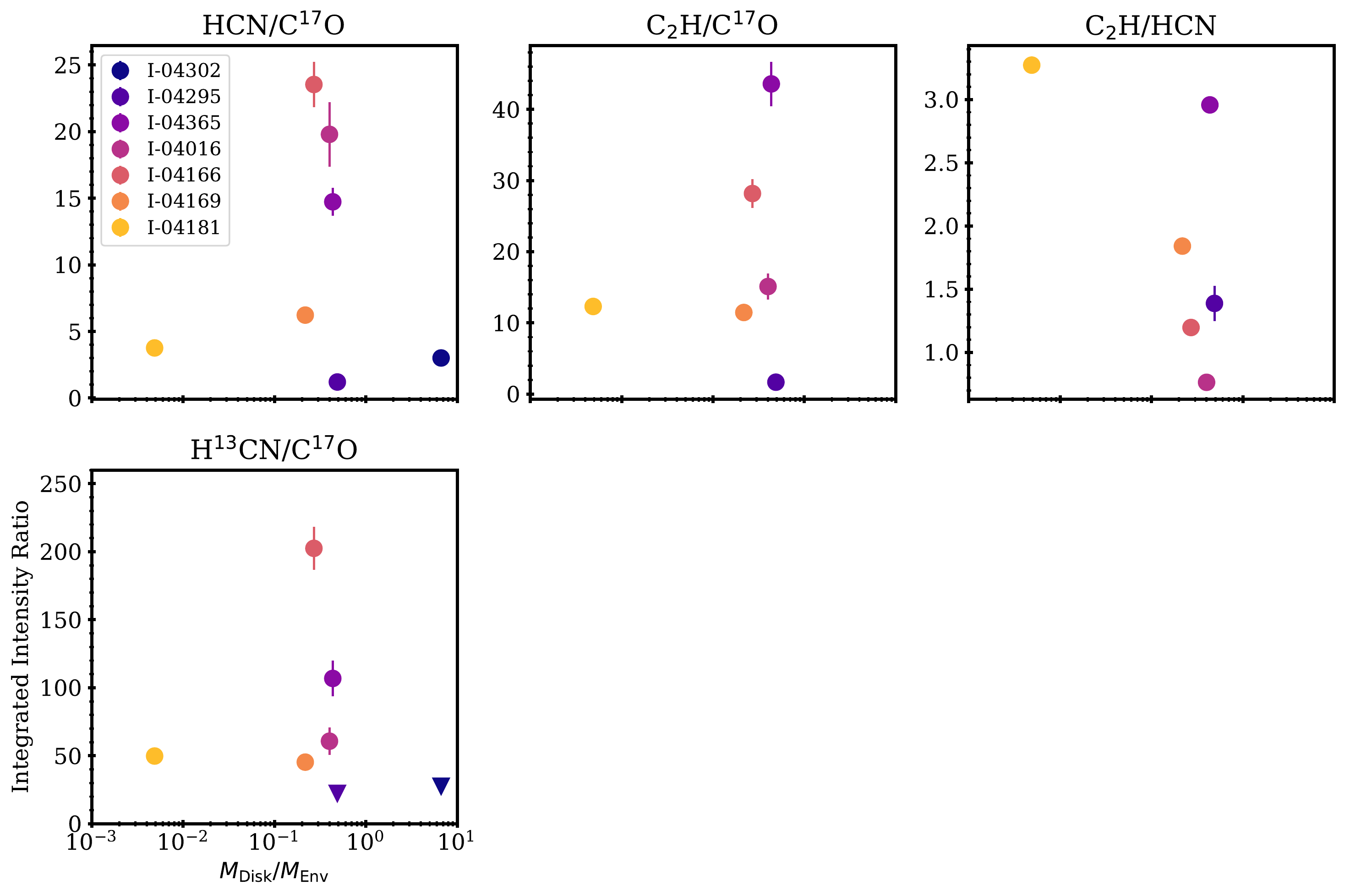}
    \caption{Integrated line flux ratios of \ce{C2H}/C$^{17}$O, HCN/C$^{17}$O, \ce{C2H}/HCN, and H$^{13}$CN/C$^{17}$O  derived from the transition $J=1-0$ of each species toward each of the seven Class I sources of our sample, represented as function of the disk-to-envelope mass ratio. Error bars are indicated by the vertical line segments. Upper limits are represented by the downward triangles. 
 }
    \label{fig:CON_ratios}
\end{figure*}

CO, C$_2$H, and HCN have been proposed to trace the gas-phase C/N/O ratios \citep[e.g.,][]{du2015,bergin2016,cleeves2018,bergner2019}. 
Figure~\ref{fig:CON_ratios} shows the integrated $1-0$ line flux ratios of HCN/C$^{17}$O, \ce{C2H}/C$^{17}$O, \ce{C2H}/HCN, and H$^{13}$CN/C$^{17}$O, as functions of the disk-to-envelope mass ratio. There is no clear monotonic trend for any of the ratios, suggestive of a lack of consistent change in C/N/O ratios with disk-to-envelope mass ratio, a proxy for Class I evolutionary stage. We note, however, that this is preliminary since HCN is found to be highly optically thick in 4/7 sources and marginally thick in the three remaining sources (see Appendix~\ref{app:opacities}). We expect that this is also the case for C$_2$H in most of the sources. Moreover, different sources likely present different excitation environments. The tentative negative correlation between C$_2$H/HCN and disk-to-envelope mass ratio likely reflects a combined changing excitation and chemical environment as Class I disks evolve and requires spatially resolved observations to interpret further. If the trend is confirmed with spatially resolved optically thin tracers, part of the chemical explanation could suggest a decreasing C/N fraction due to C sequestration into grain mantles. However, other
interpretations would also be possible, since \ce{C2H} has been shown to be very sensitive to UV environment in disks \citep{bergin2016,bergner2019,miotello2019}. For instance, an increase of atomic C and \ce{C+} in the gas phase of the sources where radiation can propagate more easily could also explain this tentative trend.

\subsubsection{Deuterium Fractionation}
\label{subsec:D_H_ratio}

Deuterium enrichment of molecules is intimately linked to their environments \citep[e.g.,][]{lubowich2000,dalgarno1984,roberts2003,willacy2007,mumma2011,huang2017,aikawa2018}. Constraints on deuterium fractionation are thus commonly used as a diagnostic of chemical inheritance versus chemical reset along the evolutionary path from molecular clouds to planetary systems \citep[we refer the reader to the reviews from][for more exhaustive overviews]{caselli2012,ceccarelli2014}. 

\begin{figure*}
    \includegraphics[scale=0.48]{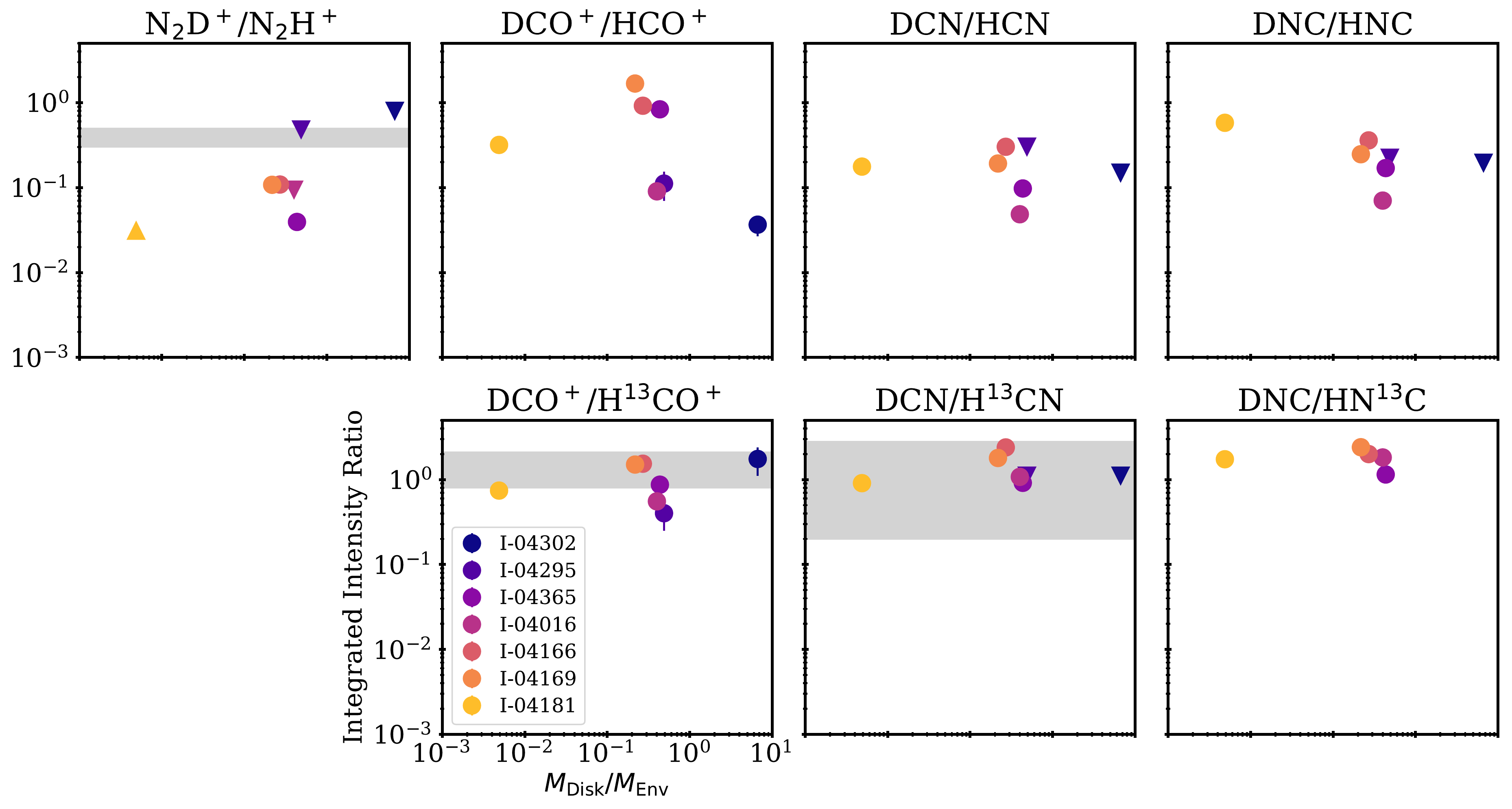}
    \caption{Integrated line flux ratios of \ce{N2D+}/\ce{N2H+}, \ce{DCO+}/\ce{HCO+}, DCN/HCN,  DNC/HNC, \ce{DCO+}/\ce{H$^{13}$CO+}, DCN/H$^{13}$CN and DNC/HN$^{13}$C $1-0$ derived for the transition $J=1-0$ in each of the seven Class I sources of our sample, sorted by decreasing disk-to-envelope mass ratios. The gray shaded regions represent the line flux ratios derived toward Class II disks from \cite{huang2015} and \cite{salinas2017} for the  \ce{N2D+}/\ce{N2H+} and from \cite{huang2017} for the \ce{DCO+}/\ce{H$^{13}$CO+} and DCN/H$^{13}$CN ratios. Upper limits are represented by the downward triangles and lower limits by the upward triangles. Error bars are indicated by the vertical line segments.}
    \label{fig:D_H_ratios}
\end{figure*}

Figure~\ref{fig:D_H_ratios} shows the line ratios \ce{N2D+}/\ce{N2H+}, \ce{DCO+}/\ce{HCO+}, DCN/HCN,  DNC/HNC, \ce{DCO+}/\ce{H$^{13}$CO+}, DCN/H$^{13}$CN, and DNC/HN$^{13}$C across our source sample. All ratios indicate highly deuterated gas, but this is in some part an opacity effect. Using the resolved hyperfine patterns of  the HCN $1-0$ line and the \ce{HCO+} $1-0$ isotopologues, we find that both, HCN and \ce{HCO+} $1-0$, are optically thick toward most of the targeted sources (see Appendix~\ref{app:opacities}), and we suspect this is also the case for the HNC $1-0$ line. However, \ce{N2H+}, H$^{13}$CN, and \ce{HCO+} are found to be optically thin in all sources where they are detected (see Appendix~\ref{app:opacities}) and can be used to directly trace the D/H ratio.

Focusing on the optically thin line ratios, we see between a factor of two and five difference for the different species. Assuming that the $^{12}$C/$^{13}$C ratio remains unchanged among our source sample, it is striking how constant the D/H ratio is across the Class I sample. Furthermore, there is no evident trend with D/H ratio and disk-to-envelope mass ratio. Additional observations are required to further this preliminary D-fractionation study.

Comparison with Class II disk observations are represented in Fig.~\ref{fig:D_H_ratios} for the $3-2$ ratios of \ce{DCO+}/\ce{H$^{13}$CO+} and DCN/H$^{13}$CN derived in a sample of six Class II disks by \cite{huang2017}, and for the  $3-2$ line ratio \ce{N2D+}/\ce{N2H+} derived toward the disks surrounding the T Tauri star AS~209 \citep{huang2015} and the Herbig-Ae star HD~163296 \citep{salinas2017}. Despite the facts that we are using a different set of transitions and that the DCN/H$^{13}$CN ratio spanning over one order magnitude in Class II disks is loosely constrained, this preliminary comparison already shows that the \ce{DCO+}/\ce{H$^{13}$CO+} and DCN/H$^{13}$CN ratios are in the same order of magnitude from Class I to Class II sources. Especially, in Class II disks the \ce{DCO+}/\ce{H$^{13}$CO+} only spans over a factor $\sim3$ and the Class I \ce{DCO+}/\ce{H$^{13}$CO+} ratios are all within a factor of 2 of this range. However, additional observations of these ratios in both Class I and Class II sources are required to further this comparison. The \ce{N2D+}/\ce{N2H+} $1-0$ line ratios derived in our Class I source sample are lower by factors of a few to almost an order of magnitude compared to the \ce{N2D+}/\ce{N2H+} $3-2$ ratios derived in Class II disks. However, 
there is a detection bias since \ce{N2D+} has been difficult to detect in Class II disks, so any \ce{N2D+}/\ce{N2H+} ratios reported in Class II disks are going to necessarily be high. 

\subsubsection{$^{14}$N/$^{15}$N Ratio In HCN}
\label{subsec:14N_15N_ratio}

In un-shielded gas, such as disk atmospheres, selective photo-dissociation of $^{14}$N$^{15}$N, whose opacity is lower than that of \ce{N2}, may result in $^{15}$N fractionation \citep[e.g.,][]{clayton2002,liang2007,heays2014}. HCN/HC$^{15}$N is a proposed probe of this process, and could potentially be linked to the UV-field strength of its environments. Chemical $^{15}$N fractionation of HCN has also long been supposed to be favored at low temperature ($<20$~K) due to the positive difference in zero-point energies between the $^{14}$N  and $^{15}$N isotopologues,  through:
\begin{equation}
    \ce{HC^{14}NH+ + ^{15}N -> HC^{15}NH+ + ^{14}N},
\label{eq:15n_frac}
\end{equation}
followed by the dissociative recombination with electron \ce{HC^{15}NH+ + e- -> HC^{15}N + H}  \citep{terzieva2000}. But, subsequent work demonstrated that the chemical fractionation of Eq~(\ref{eq:15n_frac}) is unlikely to happen at low temperatures due to an important activation barrier \citep{roueff2015}, ruling out a temperature-driven chemical fractionation of HCN. However, \cite{wirstrom2018} pointed out that the updated reaction rates from \cite{roueff2015} should be taken with caution, as they were only calculated for one collision angle. Further investigations of the low temperature $^{15}$N-fractionation pathway are still required to completely rule out the temperature-driven chemical differentiation of HCN.

\begin{figure}
    \centering
    \includegraphics[scale=0.75]{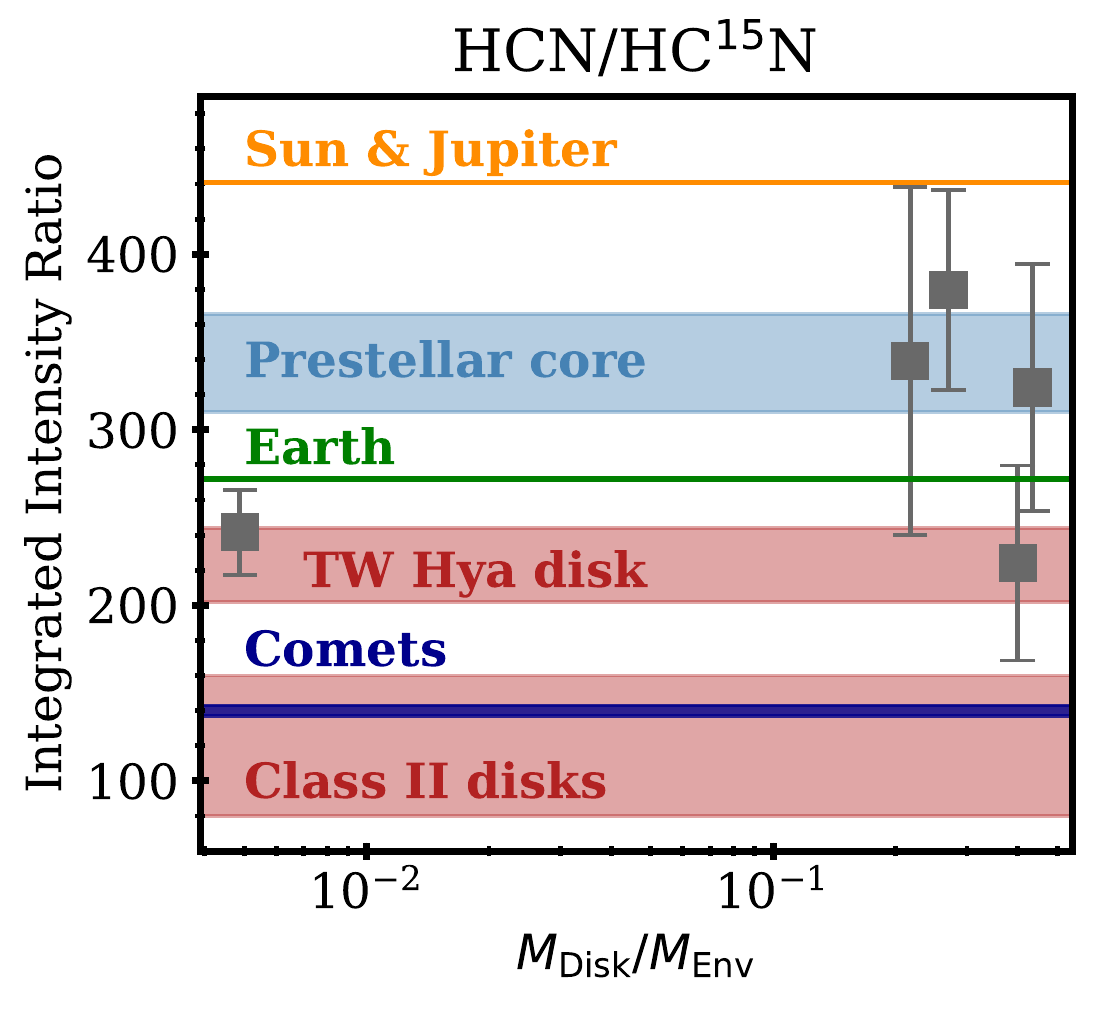}
    \caption{Integrated line flux ratio of HCN/HC$^{15}$N $J=1-0$ represented as function of the increasing disk-to-envelope mass ratio toward 5/7 of the Class I targeted sources where HC$^{15}$N is at least tentatively detected. These ratios were derived from the H$^{13}$CN/HC$^{15}$N $J=1-0$ ratio, assuming the local ISM isotopic ratio of $^{12}$C/$^{13}$C of 68 (see text \S~\ref{subsec:14N_15N_ratio}). The horizontal lines and rectangles represent the Solar (orange), Prestellar core (light blue), Earth (green) Class II disks (red), and Comets (dark blue) ratios for references. Error bars are indicated by the vertical line segments.}
    \label{fig:14N_15N_ratio}
\end{figure}

Figure~\ref{fig:14N_15N_ratio} shows the HCN/HC$^{15}$N ratios derived toward the five sources of our sample where H$^{13}$CN and HC$^{15}$N are at least tentatively detected, using the integrated line flux ratio of H$^{13}$CN/HC$^{15}$N $J=1-0$ and assuming a constant $^{13}$C/$^{12}$C isotopic ratio of $68\pm 15$ \citep{milam2005, manfroid2009,asplund2009}, and how they compare with HCN/HC$^{15}$N ratios from other kinds of sources. The Class I disks span the ratio between the TW~Hya disk \citep{hilyblant2019} and prestellar cores \citep{magalhaes2018}. The  HCN/HC$^{15}$N Class I measurements are higher by factors of $\sim1.5-4$ than the ratios measured in a sample of five Class II disks \citep[spanning from 80 to 160,][]{guzman2017} and in comets \citep[with an average ratio of $140\pm3$,][see also the compilation made in Fig. A1 of \cite{hilyblant2017}]{jehin2009,mumma2011,bockelee-morvan2015,shinnaka2016}. However, the ratio may be somewhat overestimated in I-04016 and I-04169 where HC$^{15}$N was only tentatively detected (see Table~\ref{tab:intensities}). 

Moreover, these are all source-averaged ratios. New higher-resolution measurements of the HCN/HC$^{15}$N toward the TW~Hya disk found that the isotopic HCN ratio varies from $121\pm11$ to $339\pm28$ from inner to outer radii \citep{hilyblant2019}. These new results suggest that the isotopic fractionation is occurring within the Class II disk stage resulting in $^{15}$N enrichment from outer to inner disk regions where comets are supposed to originate. This ratio may then also be a good tracer of how exposed the disk is to UV radiation in the younger Class I disks, but high-resolution Class I disk observations are needed to confirm this.

\subsubsection{CN/HCN: Probe Of The UV-field Strength?}
\label{subsec:CN_HCN_ratio}

\begin{figure}
    \centering
    \includegraphics[scale=0.75]{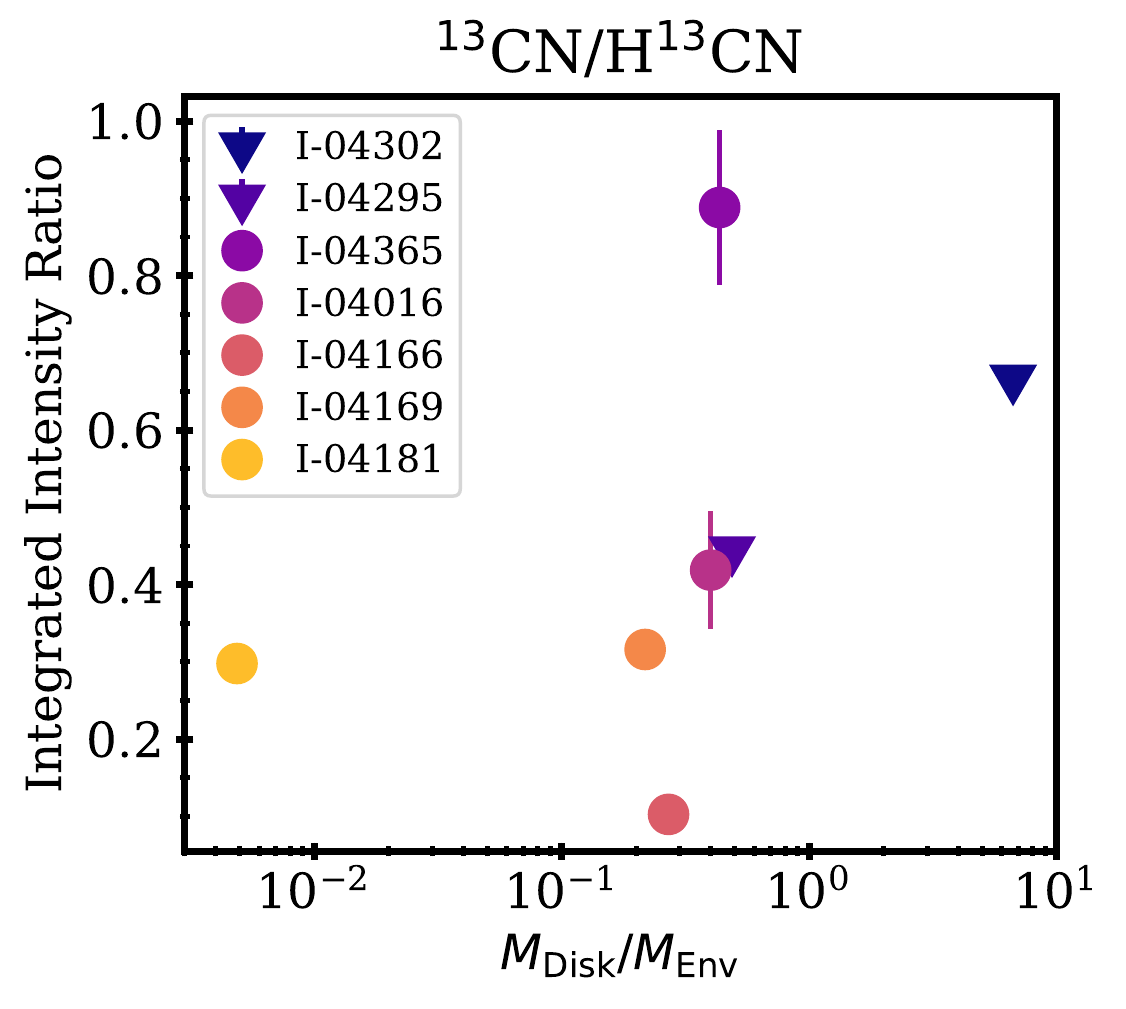}
    \caption{Integrated line flux ratios of the $^{13}$CN/H$^{13}$CN $J=1-0$ as function of the increasing disk-to-envelope mass ratio of our Class I source sample. Upper limits are represented by the downward triangles. Error bars are indicated by the vertical line segments.}
    \label{fig:CN_HCN}
\end{figure}

The CN/HCN ratio is known to be highly correlated with UV-field strength of the environment it probes \citep[e.g.,][]{schilke1992,fuente2003,riaz2018,bulblitz2019}. In protostellar sources in particular, it has been shown to probe the UV flux from the accretion regions of the embedded protostar and its forming circumstellar disk \citep[e.g.,][]{stauber2007}. In these regions where the 
stellar FUV flux is higher than the interstellar UV radiation field, the CN/HCN ratio can be enhanced through photodissociation of HCN driven by Ly$\alpha$ radiation from the central star \citep{bergin2003}, as mentioned earlier (\S~\ref{subsec:line_flux_overview}). CN/HCN line ratios $\gtrsim 10$ have been reported in Class II T Tauri \citep[DM~Tau;][]{dutrey1997} and Herbig Ae disks \citep[MWC~480 and HD~163296;][]{thi2004,chapillon2012b}. In Class II disks the disk averaged CN/HCN does not vary much from one disk to another \citep{oberg2011a}, however, but it does vary within disks \citep{guzman2015disk}. 

Figure~\ref{fig:CN_HCN} shows the integrated line flux ratio of $^{13}$CN/H$^{13}$CN $J=1-0$ of our Class I source sample, which varies by almost one order of magnitude across our source sample. The source with the lowest $^{13}$CN/H$^{13}$CN ratios, is the one that we previously identified as the coldest of our sample due to the high intensity of deuterated species and to its high HNC/HCN ratio. The highest $^{13}$CN/H$^{13}$CN ratio is found in  I-04365.

Based on the $^{13}$CN/H$^{13}$CN $J=1-0$ line ratio as a proxy for the CN/HCN abundance ratio and assuming that the latter is a robust UV-field strength probe, I-04365 should be the most exposed sources to UV radiation. 
However, the H$^{13}$CN/HC$^{15}$N ratio, also proposed as a probe for the UV-field strength (see Section~\ref{subsec:14N_15N_ratio}), is found to be the lowest in I-04016. 
I-04016 and I-04365 are the two most luminous sources of our sample (see Table~\ref{tab:sources}). Thus, while the results found with the $^{13}$CN/H$^{13}$CN and H$^{13}$CN/HC$^{15}$N ratios tend to be consistent, the differences found might indicate that either one of these ratios also depends on other parameters; or that each of these ratios trace different layers or components of the Class I source probed. Additional observations are required to test these scenarios.

\subsubsection{HNC/HCN: A Probe Of Temperature?}
\label{subsec:HCN_HNC_ratio}

\begin{figure}
    \centering
    \includegraphics[scale=0.75]{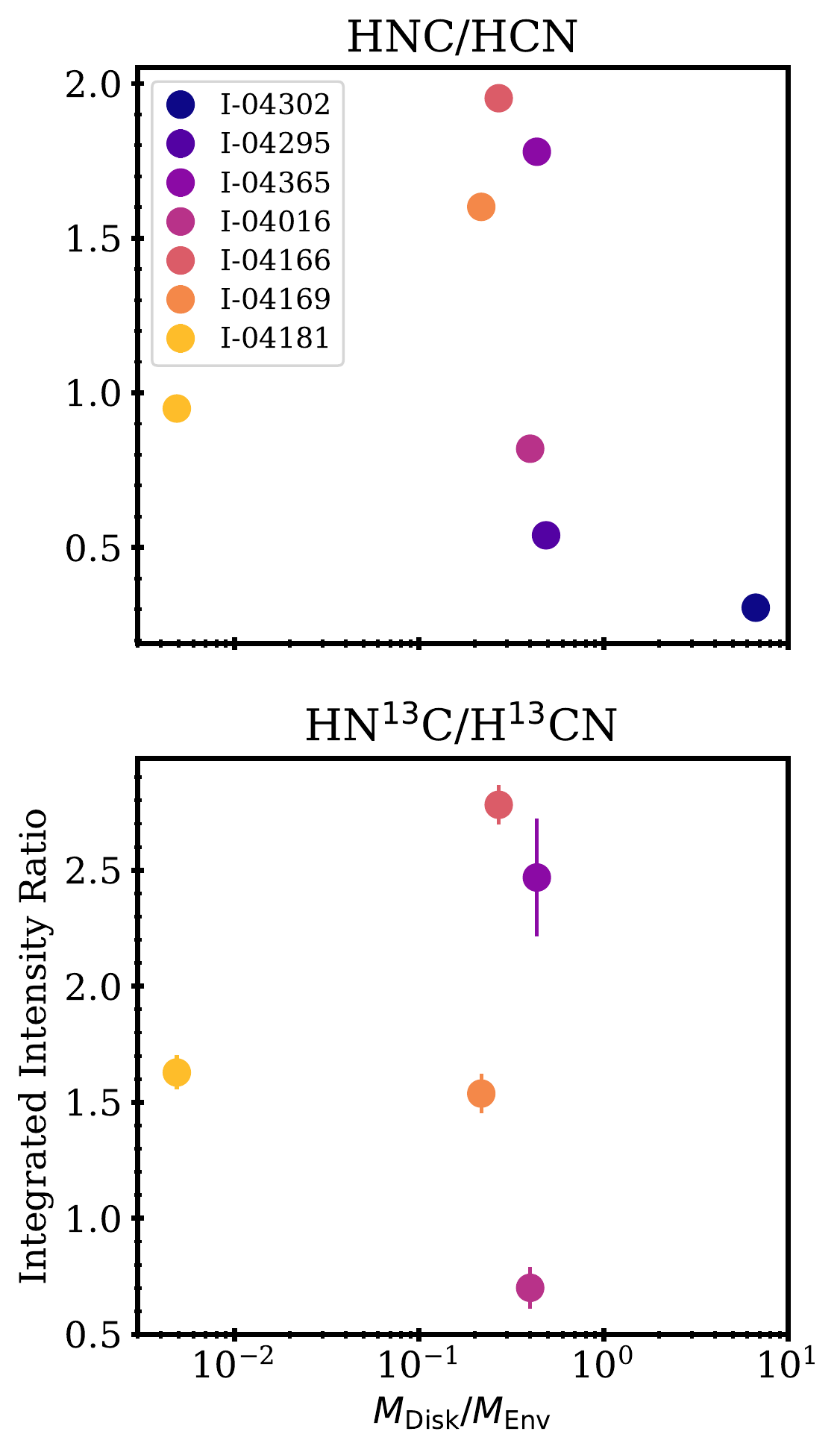}
    \caption{Integrated line flux ratios of the HNC/HCN (top) and HN$^{13}$C/H$^{13}$CN (bottom) $J=1-0$ as function of the increasing disk-to-envelope mass ratio of our Class I source sample. Upper limits are represented by the downward triangles. Error bars are indicated by the vertical line segments.}
    \label{fig:HNC_HCN}
\end{figure}

The HNC/HCN $J=1-0$ line ratio is suggested to be a  direct probe of the gas kinetic temperature, such that a higher ratio probes colder material \citep[e.g.,][]{baan2008,hacar2019}.
In Fig.~\ref{fig:HNC_HCN} we show the HNC/HCN and HN$^{13}$C/H$^{13}$CN $J=1-0$ line ratios. When comparing the top and bottom panels of this figure, we see that the order of the two ratios from one source to another is generally replicated. The lowest ratios are found in the sources with the highest disk-to-envelope mass ratios, suggesting they are overall warmest. Among the envelope dominated sources, there is no trend with disk-to-envelope mass ratio. However, this is not surprising since source luminosity should also play a large role for setting the temperature. In fact, we do see the highest HNC/HCN ratio toward the lowest luminosity source, I-04166,  which is also the coldest source of our sample (see Table~\ref{tab:sources}). Interestingly, the H$^{13}$CN/HC$^{15}$N is also found to be the highest in this source, highlighting the fact that the latter is indeed related to the luminosity of the source and that $^{15}$N fractionation of HCN is less efficient at low temperature (see Section~\ref{subsec:14N_15N_ratio}).

\subsubsection{Sulfur-bearing Species Ratios}
\label{subsec:S_ratios}

Sulfur-bearing species ratios have long been thought to be powerful diagnostics to probe both the chemistry and the physics of their environments. For instance, the CS/SO ratio has been proposed as a probe for the elemental C/O ratio \citep[e.g][]{swade1989,bergin1997,fuente2019}.  
Sulfur-bearing ratios are also thought to be good chemical clocks (e.g., SO/\ce{SO2}, CS/OCS, SO/OCS), because of their rapid chemical evolution with gas temperature and density \citep[e.g.,][]{charnley1997,hatchell1998,viti2004,herpin2009,wakelam2011,vidal2018}. This idea came up from hot core/corino theory, with the hypothesis that the bulk of the sulfur reservoir is locked onto grain mantles in previous evolutionary stage and then released in the gas phase during the hot core/corino formation under the form of \ce{H2S}, SO and atomic S. This is supposed to initiate an active S-chemistry, where \ce{SO2} is a direct product of SO, through the radiative association of O and SO and the neutral-neutral reaction \ce{OH +SO -> SO2 + H}, making the SO/\ce{SO2} ratio an ideal chemical clock for such astrophysical objects. In the gas phase, OCS is mainly formed through the radiative association of S and CO, and one of the formation pathways for CS is the reaction \ce{SO + C}, justifying that CS/OCS and SO/OCS were also claimed as additional chemical evolutionary clocks. Finally, finding out whether sulfur is reduced or oxidized in planet-forming environments is interesting for origins of life chemistry \citep{patel2015}.

\begin{figure*}
    \centering
    \includegraphics[scale=0.43]{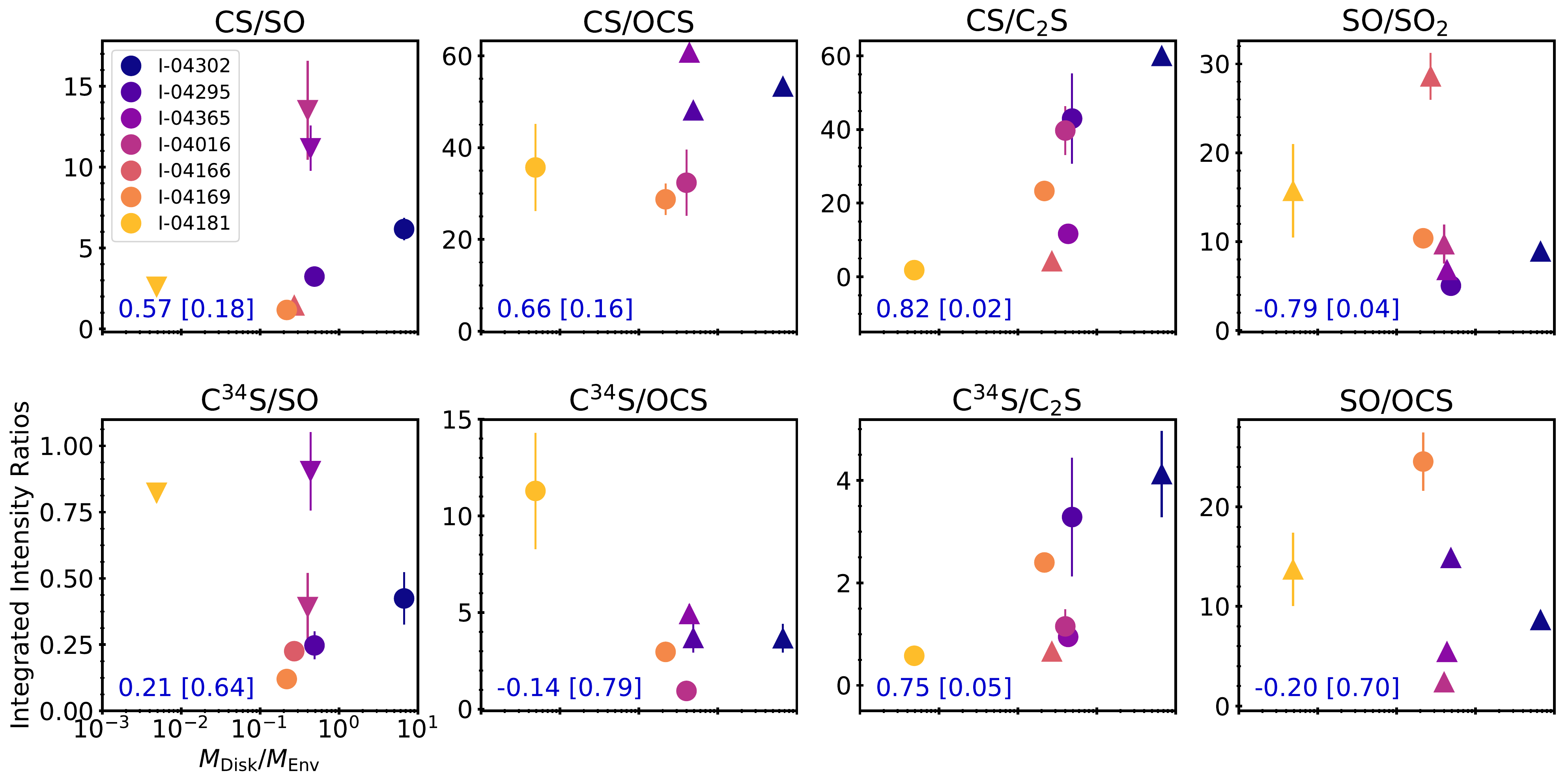}
    \caption{Integrated line flux ratios of CS/SO, CS/OCS, CS/\ce{C2S}, SO/\ce{SO2}, and SO/OCS, with corresponding  C$^{34}$S ratios on the bottom row panels, derived toward our Class I source sample. These ratios are represented as function of the increasing disk-to-envelope mass ratio. Upper limits are indicated by the downward triangles and lower limits by the upward triangles. Error bars are indicated by the vertical line segments. Spearman correlation coefficients are indicated in the lower left corner of each panel, with corresponding p-value in brackets. These statistical parameters are described in \S~\ref{subsec:line_flux_overview}. Note that correlation calculations include upper limits.}
    \label{fig:S_ratios}
\end{figure*}

Figure~\ref{fig:S_ratios} shows the S-bearing integrated line flux ratios of CS/SO, CS/OCS, CS/\ce{C2S}, SO/OCS, and SO/\ce{SO2} 
with corresponding  C$^{34}$S ratios as functions of increasing disk-to-envelope mass ratio across the Class I sample. 
These ratios are also found to vary by several orders of magnitude from one source to another, but only CS/\ce{C2S} (and C$^{34}$S/\ce{C2S}) show a clear positive correlation with increasing disk-to-envelope mass ratio, respectively. This is {\it a priori} counter-intuitive because, for instance, \ce{C2S} can be produced from the reaction \ce{CS + CH -> C2S + H} in the gas phase (see the KIDA database\footnote{http://kida.obs.u-bordeaux1.fr/}). Hence, one would rather expect an increase in carbon chains with evolution. However, it should be noted that the chemistry of \ce{C2S} has remained poorly studied experimentally and theoretically so far.

\begin{figure*}
    \centering
    \includegraphics[scale=0.4]{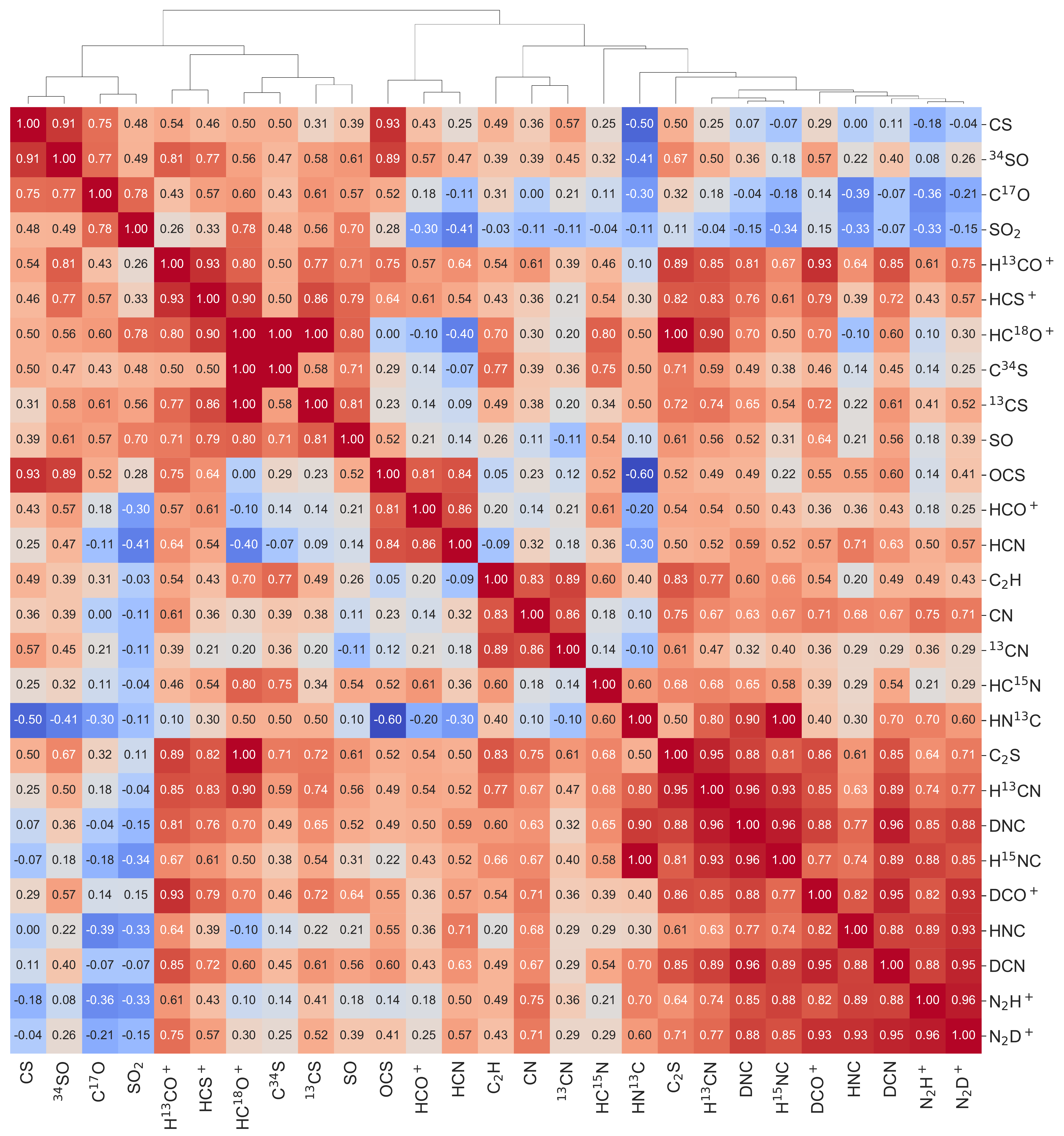}
    \caption{Cluster correlation matrix plot for all the 27 species presented in this work, based on Spearman's correlation coefficients indicated in each box. This statistical parameter is described in \S~\ref{subsec:line_flux_overview}. The dendrogram illustrates the arrangement produced by the clustering analysis. The redder the color is, the more positive the correlation is and, similarly, the bluer the color is, the higher the anti-correlation is. Note that correlation calculations include upper limits.}
    \label{fig:full_cluster}
\end{figure*}

\section{discussion}
\label{sec:discussion}
\subsection{Chemical Differentiation Within Our Sample}

In Section~\ref{sec:results}, we explored specific line emission patterns within our sample. In this sub-section we investigate global patterns of chemical differentiation.
Fig.~\ref{fig:full_cluster} shows a cluster correlation matrix of all sources and line fluxes presented in this work, based on Spearman's correlation coefficients. This correlation matrix method produces a table containing the correlation coefficients in between each species, allowing us to simultaneously assess their interdependence. The clustering is then used to arrange and group the species by correlation coefficients, as illustrated by the dendrogram in the figure.

This global statistical representation highlights several clusters of correlated molecules with two or more sub-clusters each. At the bottom right of the plot lies the biggest cluster consisting of \ce{N2D+}, \ce{N2H+}, DCN, HNC, {\ce{DCO+}, H$^{15}$NC, DNC, H$^{13}$CN, \ce{C2S}, and HN$^{13}$C, which are readily identified with dense and cold gas, since these species are known to have high critical densities \citep[i.e., tracing dense gas,][]{shirley2015}, and were observed in their 1-0 ground-state transition in our survey (i.e., tracing cold gas). Furthermore, deuterated species and especially \ce{N2D+} are characteristics of cold gas for chemical reasons.}
Bright HNC emission is also associated with cold gas, while HCN isotopologues are more of dense gas tracers. But, the fact that they show up in the same cluster may suggest that sources that are cold are also more likely to have large reservoirs of dense gas. In either case, the first differentiating factor among Class I sources affecting their temperature is whether or not they have a large cool gas reservoir. 

At the top left of the figure lies the second biggest cluster, overall less strongly correlated than the previous one, but composed of several highly correlated sub-clusters. It contains most of the S-bearing species but OCS and \ce{C2S}.  
With its abundance in oxygenated S-bearing species, this molecular cluster is probably characteristic of shock induced chemistry tracers, which appear to be closely correlated with CS and isotopologues, C$^{17}$O, and with ions such as \ce{H$^{13}$CO+} and \ce{HC$^{18}$O$^+$}. The coupling of these last species with shock tracers could either be due to an evolutionary effect, or to the fact that sources with shocks may also have more ionized gas and/or more CO in the gas-phase. C$^{17}$O (our proxy of CO in the gas phase, see Section~\ref{subsec:line_det}) is indeed found to be the brightest in I-04169, where most S-bearing species and \ce{H$^{13}$CO+} are also the brightest.

Finally, two smaller clusters lie in the center of the matrix: {\it(i)} a first one made of {\it(i)} \ce{HCO+}, OCS, and HCN; and {\it(ii)} another one made of CN, \ce{^{13}CN}, and \ce{C2H}. Interestingly, the first molecular cluster is however barely correlated with the fainter isotopologues of \ce{HCO+} and HCN, which suggests that the correlations of these molecules are dominated by excitation and opacity effects rather than chemical relationships. The second cluster contains UV-driven dense gas tracers, with CN and \ce{C2H} being commonly observed in UV-enhanced environments \citep[e.g.,][]{kastner2015,bergin2016,cazzoletti2019,vanterwisga2019,bergner2019,miotello2019}.

\subsection{\ce{SO2} Versus C$^{17}$O In Class I Sources}

A recent study of twelve low-mass Class I sources located in the Ophiuchus molecular cloud shed light on the chemical differentiation between \ce{SO2} and \ce{C$^{17}$O} across their source sample \citep{delavillarmois2019}. \ce{SO2} emission was found to be the brightest toward the most luminous sources, whereas the opposite was observed for \ce{C$^{17}$O}. The fact that these two molecules are not tracing similar environments is not surprising, since these two molecules are known to trace different physical conditions, where the former is often associated with outflows and shocked gas \citep{jorgensen2004,podio2015}, whereas the latter is rather tracing denser and colder gas \citep{caselli1999,tafalla2002,redman2002}. However, \cite{delavillarmois2019} propose that in Class I sources, the ratio of these two molecules could provide an interesting evolution diagnostic, assuming that the luminosity of the source is linked to the evolutionary stage of the source.

\begin{figure}
    \centering
    \includegraphics[scale=0.75]{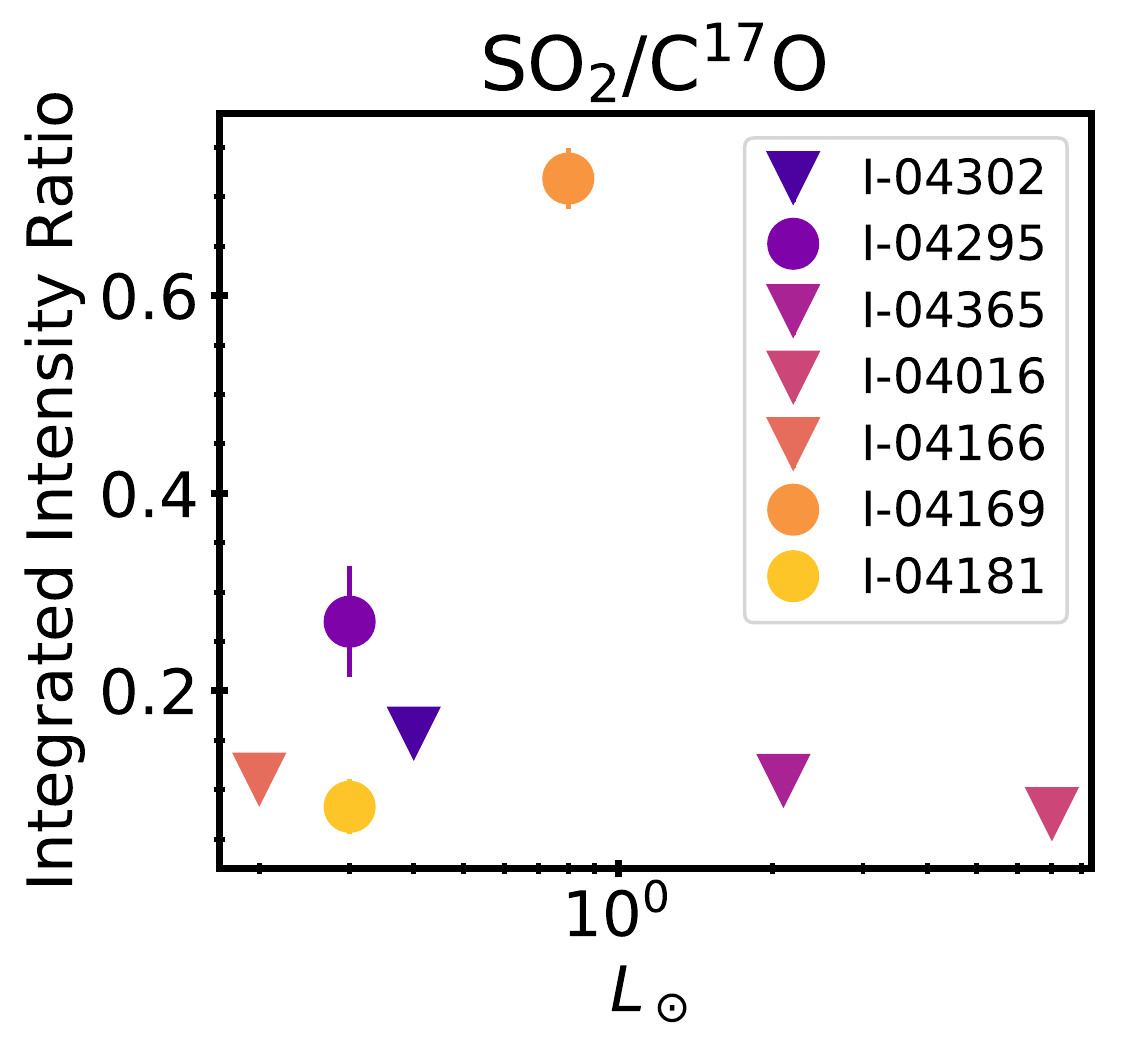} 
     \caption{\ce{SO2}/\ce{C$^{17}$O} integrated line flux ratio as function of the source luminosity across our Class I sample. Upper limits are represented by the  downward triangles.}
    \label{fig:SO2_C17O}
\end{figure}

Figure.~\ref{fig:SO2_C17O} represents the integrated line flux ratios \ce{SO2}/\ce{C$^{17}$O} across our Class I sample as function of the luminosity of the sources. First, we can see that in our sample the luminosity and disk-to-envelope mass ratio are not correlated and luminosity alone is thus not likely a good evolutionary indicator here. For the sources where the \ce{SO2} line we targeted here is detected (i.e., 4/7 in our sample), there is a noticeable increase in the \ce{SO2}/\ce{C$^{17}$O} ratio with luminosity, but the highest luminosity sources present informative upper limits on this ratio which dramatically breaks the trend.

\subsection{Class 0 versus Class I versus Class II chemistry}

As opposed to Class I systems, many more spectral surveys have been performed to study the chemistry of the earliest stages of star  formation, i.e. Class 0 objects -- e.g., TIMASSS \citep{caux2011}, PILS \citep{jorgensen2016}, ASAI \citep{lefloch2018}, SOLIS (Ceccarelli et al. 2017) --  and of late planet-forming  disks,  i.e. Class II objects -- DISCS \citep{oberg2010,oberg2011a}, CID \citep{guilloteau2016}. The question that is still pending is to determine if the chemical composition of Class I objects is more Class 0-like, believed to be associated with \ce{CH3OH} and long carbon chains \citep[e.g.,][]{bergner2017,law2018} or Class II-like, dominated by small molecules and an O-poor organic chemistry \citep[e.g.,][]{oberg2011a,guilloteau2016}. Although the non-detection of larger molecules in Class II objects could be due to a detection bias, as the column densities typically decrease from Class 0 to Class II stages, non-oxygenated complex molecules, such as \ce{HC3N} and \ce{CH3CN}, are commonly observed in Class II disks \citep[e.g.,][]{bergner2018} as compared to O-containing complex molecules, such as \ce{CH3OH}, therefore highly favoring that disk organic chemistry is indeed O-poor \citep[e.g.,][]{legal2019b}.

In the present work, we focused on small ($N_{\rm{atoms}}\leq 3$) molecules, with the detections of a wide variety of C, N, O, and S carriers - namely CO, \ce{HCO+}, HCN, HNC, CN, \ce{N2H+}, \ce{C2H}, CS, SO, \ce{HCS+}, \ce{C2S}, \ce{SO2}, OCS - and some of their D, $^{13}$C, $^{15}$N, $^{18}$O, $^{17}$O, and $^{34}$S isotopologues. This set of detections is thus already providing interesting information regarding the fact that Class I systems are richer than Class II objects in small molecules, since all the Class II disk species detected so far are also present in Class I stage and that, in addition, oxygen-bearing species are also detected in Class I stage, such as SO, \ce{SO2}, and OCS. In particular, we note that SO is systematically detected in all the sources of the Class I sample targeted here, contrary to Class II disks where it has only been reported toward a few young disks presenting active accretion signs \citep{fuente2010,guilloteau2016,pacheco2016,booth2018}. 

However, interferometric observations are needed to distinguish which source components are traced by each of these species. This would be, in particular, beneficial to better understand the lack of O-bearing species in Class II disks, such as the lack of oxygenated S-bearing species reported so far \citep[e.g.,][]{semenov2018,legal2019a}. As for the comparison with Class 0 objects, complex organic detections from the same Class I survey will be presented in a forthcoming study and should thus allow to identify if long carbon chain and/or complex organic molecules are still present in Class I stage and in which quantities and proportions as compared to small organics. 

\section{Summary and Conclusion}
\label{sec:conclusion}
We presented a new single-dish 3mm spectral line survey carried out with the IRAM-30m telescope toward a sample of seven Class I YSOs. Our main findings are summarized below:
\begin{enumerate}
\item Our 3mm survey found that Class I YSOs are molecule-rich, considering the large number of spectral lines we found (see Fig.~\ref{fig:all_sces_full_3mm}). We report the detections of twenty-seven spectral lines characterizing the main small ($N_{\rm{atoms}}\leq 3$) molecules of our survey - namely CO, \ce{HCO+}, HCN, HNC, CN, \ce{N2H+}, \ce{C2H}, CS, SO, \ce{HCS+}, \ce{C2S}, \ce{SO2}, OCS, including some of their D, $^{13}$C, $^{15}$N, $^{18}$O, $^{17}$O, and $^{34}$S isotopologues.

\item For most species, integrated line intensities vary by 1-2 orders of magnitude across the source sample,  even  within  the  central  narrow  range  of  disk-to-envelope  mass  ratios  varying  from  0.2  to  0.5. There is a general negative trend in line intensity with increasing disk-to-envelope ratio, used as a proxy for the Class I evolutionary stage, but the more salient point is the large intensity variations found among sources with similar total masses.

\item Line flux ratios proposed to trace C/N/O gas-phase ratios, do not show clear monotonic trends with disk-to-envelope mass ratio. A tentative negative correlation between \ce{C2H}/HCN and the disk-to-envelope mass ratio may indicate a decreasing C/N fraction due to C sequestration into grain mantles, although other scenarios, e.g. driven by UV-penetration, might also be possible. 

\item Line flux ratios that probe deuterium fractionation are rather constant across the Class I sample, which indicates similar thermal histories and/or current thermal structure toward Class I objects. Higher angular and spectral resolution observations are required to corroborate these conclusions.

\item Comparing the HCN/HC$^{15}$N measured in our Class I sample with ratios from other astrophysical environments, we found that the Class I ratios span between prestellar core and Class II disk values. This might either be indicative of the span in evolution of our sample, if the isotopic ratio is inherited; or of the span in physical properties of our sample, if the HCN/HC$^{15}$N is reprocessed {\it in situ}, since this ratio is likely to be strongly altered with UV radiation in un-shielded environments where the visual extinction lies within 1 to 3 magnitudes \citep[e.g.,][]{heays2014}.

\item The CN/HCN ratio, proposed as a UV-field strength diagnostic, varies by  almost one order of magnitude across our source sample. Its highest values are reached in I-04365 suggesting that this source might be the most UV-illuminated, but detailed abundance modelling is needed to confirm.

\item The HNC/HCN ratio, proposed as a gas temperature proxy, is lower toward the sources with higher disk-to-envelope ratios, and the highest 
toward the source with the lowest luminosity and bolometric temperature, I-04166, where species characteristic of cold gas are the brightest.
\item We explored the use of ratios of S-bearing species -- CS/SO, CS/OCS, CS/\ce{C2S}, SO/\ce{SO2}, and SO/OCS -- as tracers of the C/O elemental ratio and chemical evolutionary stage. Only CS/\ce{C2S} and SO/\ce{SO2} show correlations with the disk-to-envelope mass ratio, which are found to be opposite. These opposite trends could indicate that these ratios probe different physical components of the Class I sources, but it could also be due to oxidized versus reduced S-chemistry. Higher angular resolution observations are required to further explored these results.

\item To illustrate the chemical differentiation within our sample, we built a cluster correlation matrix that highlights three main groups of correlated species: {\it(i)} dense cold gas tracers, {\it(ii)} shocked gas tracers, and {\it(iii)} dense ionized gas tracers - likely probing more UV-illuminated regions than the previous ones.

\item To compare our results to other Class I surveys, we focused on the \ce{SO2}/C$^{17}$O ratio proposed to be a Class I evolutionary tracer, assuming that the latter is correlated with source luminosity \citep{delavillarmois2019}. First, in our sample, the disk-to-envelope mass ratio that we proposed as a proxy for evolutionary stage is not correlated with source luminosity. Second, although there could be a tentative correlation in between this ratio and source luminosity within our sample, the fact that the ratios derived in the highest luminosity sources rely on upper limits breaks the trend. 

\item Finally, we discuss the similarities and differences we found when comparing our Class I chemical results to Class 0 and Class II chemistry. Class I sources are found to be more chemically rich and diverse than Class II sources, the latter being known to harbor an O-poor organic chemistry. Class 0 objects are usually associated with more complex molecules and long carbon chains. However, in the present work we focused on small ($N_{\rm atoms}\leq3$) molecules. The composition in bigger and more complex organic molecules within the same sample of Class I sources will be investigated in a forthcoming study.

\end{enumerate}
These results already provide important information on the chemical content and the relationship between Class I chemistry, evolutionary stage and other source properties. Since all sources are located in the same star-forming region, our analysis avoids cloud-to-cloud background variations, but as a result, it may not be representative of all regions of the sky.
Additional observations are required to extend this Class I sample and increase the statistics and demography on Class I YSO chemistry. 
Interferometric observations will be beneficial to elucidate 
the contributions of envelope and disk chemistry to the observed emission lines. 
Future work on complex organics in the same sample should also provide additional clues on the evolution and chemical content of these objects.%

\acknowledgments 
We thank the anonymous reviewer for a careful reading and constructive comments that helped us to improve the paper. This paper make use of IRAM 30M data, project ID: 014-19. IRAM is supported by INSU/CNRS (France), MPG (Germany) and IGN (Spain). RLG also thanks the IRAM staff and in particular the IRAM AOD A. Ritacco, and the two IRAM operators V. Peula and F. Damour for their support to run the telescope during the IRAM shift. J.H. acknowledges support from the National Science Foundation Graduate Research
Fellowship under grant No. DGE-1144152. C.J.L. acknowledges funding from the National Science Foundation Graduate Research Fellowship under grant No. DGE-1745303.

\software 
{\texttt{CASSIS}: http://cassis.irap.omp.eu, \texttt{CLASS/GILDAS}:  http:www.iram.fr/IRAMFR/GILDAS \citep{pety2005,gildasteam2013}, \texttt{Pandas} \citep{pandas}, \texttt{Matplotlib} \citep{matplotlib}, \texttt{NumPy} \citep{numpy_article}, \texttt{SciPy} \citep{virtanen2020}, \texttt{Seaborn} \citep{michael_waskom_2018}}.\newpage

\clearpage
\appendix
\section{Optical Depth Effects}
\label{app:opacities}
Line optical depth can be assessed when several isotopologues of the same line are detected or when the line possesses a hyperfine pattern that is resolved by the observations. 
In this appendix, we investigate the line opacity of some of the molecular species we detected possessing one of these characteristics and for which puzzling results were found. 
For instance, the brightest lines of CS, HCN, and \ce{HCO+} occur in different sources than their respective fainter isotopologues. 
Similarly, the brightest \ce{H^{13}CO+} and \ce{HC^{18}O+} lines also occur in different sources. Furthermore, it seems that the more abundant isotopologues \ce{H^{13}CO+} and CS are brighter in the source with the second smallest disk-to-envelope ratio, I-04169, whereas the rarer ones are brighter in the one with the smallest ratio, I-04181. All these results could be due to optical depth effects coming either from the lines themselves, or even from the dust continuum. Dust opacity estimate being out of the scope of this work, we focus here on line opacity study.

\subsection{Isotopologue Ratios}
\label{subsec:app_isotopic_ratio}

\begin{figure}
    \centering
    \includegraphics[scale=0.5]{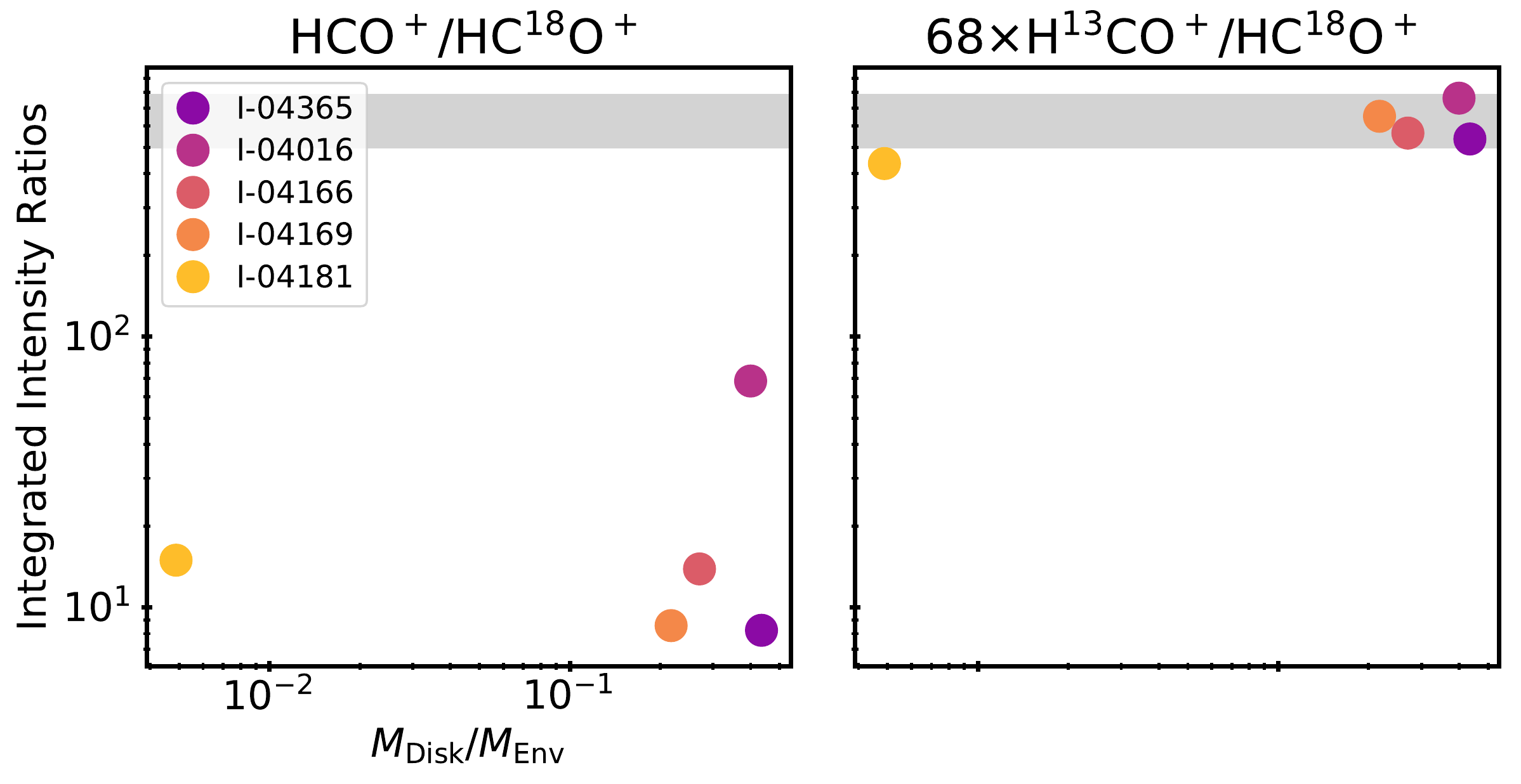}
    \caption{Integrated line flux ratios of \ce{HCO+}/HC$^{18}$O$^+$ and H$^{13}$CO$^+$/HC$^{18}$O$^+$ as function of the increasing disk-to-envelope mass ratio across the five Class I sources where HC$^{18}$O$^+$ is detected. Error bars are indicated by the vertical line segments. The H$^{13}$CO$^+$/HC$^{18}$O$^+$ ratio is multiplied by the elemental $^{12}$C/$^{13}$C$= 68\pm15$ in the local ISM \citep{milam2005,asplund2009,manfroid2009} to be compared to the \ce{HCO+}/HC$^{18}$O$^+$ ratio and to the local ISM $^{16}$O/$^{18}$O ratio (see Section~\ref{subsec:app_isotopic_ratio}), represented in gray.} 
    \label{fig:HCOp_isotopic_ratio}
\end{figure}

To investigate the opacity of the \ce{HCO+} $1-0$ and \ce{H^{13}CO+} $1-0$ lines, we derived their isotopologue line flux ratios with the \ce{HC^{18}O+} $1-0$ line, which is expected to be thin. As depicted in Fig.~\ref{fig:HCOp_isotopic_ratio}, across the seven Class I sources of our sample, \ce{HCO+}/\ce{HC^{18}O+} line ratio does not exceed 60. Across the five sources of our sample where \ce{HC^{18}O+} is detected, the \ce{H^{13}CO+}/\ce{HC^{18}O+} ratio varies from 11.2 (in I-04016) to 6.4 (in I-04181) (see Table~\ref{tab:intensities}), which leads to an \ce{HCO+}/\ce{HC^{18}O+} ratio varying from 762 (in I-04016) to 436 (in I-04181), assuming the local ISM $^{12}$C/$^{13}$C ratio of $68 \pm 15$ \citep{milam2005,asplund2009,manfroid2009}.
In the ISM, the $^{16}$O/$^{18}$O ratio is found to be $557 \pm 30$ \citep{wilson1999}, close to the Solar System value of 530 in the Solar wind \citep{mckeegan2011}, 511 $\pm$ 10 in the photosphere of the Sun \citep{ayres2013}, and $\approx500$ in comets \citep{jehin2009,bockelee-morvan2012} and meteorites \citep{lodders2003}. In the diffuse ISM, the $^{16}$O/$^{18}$O ratio is found to be a bit higher, with a value of $672 \pm 110$ derived from the \ce{HCO+}/\ce{HC^{18}O+} ratio \citep{lucas1998}. Hence, these results confirm that \ce{HCO+} $1-0$ is highly thick. On the contrary, the \ce{H^{13}CO+} $1-0$ line seems to be rather thin except in one source, I-04181. 
However, variation in the $^{12}$C/$^{13}$C and/or $^{16}$O/$^{18}$O isotopic ratios in Class I YSOs cannot be excluded. Higher spatial resolution observations are required to disentangle between these two scenarios, to determine whether the lines are tracing the same regions and physical components of the sources.

\begin{figure}
    \centering
    \includegraphics[scale=0.5]{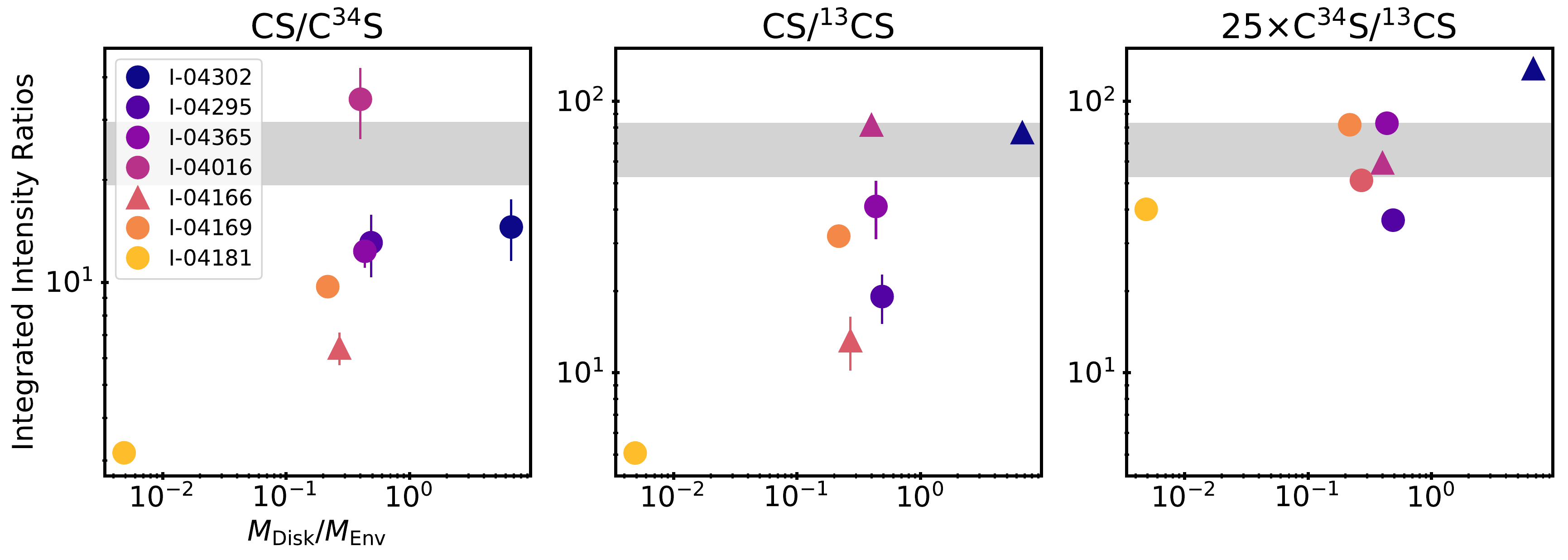}
    \caption{Integrated line flux ratios of CS/C$^{34}$S and CS/$^{13}$CS as function of the increasing disk-to-envelope mass ratio across the seven Class I sources targeted in this work. Lower limits are indicated by the upward triangles. Error bars are indicated by the vertical line segments. The elemental $^{32}$S/$^{34}$S$= 24.4\pm5.0$ in the vicinity of the Sun \citep{chin1996} and $^{12}$C/$^{13}$C$= 68\pm15$ in the local ISM \citep{milam2005,asplund2009,manfroid2009} are represented in gray.}
    \label{fig:CS_isotopic_ratio}
\end{figure}

Similarly, in Fig.~\ref{fig:CS_isotopic_ratio} we compare the CS/C$^{34}$S, CS/$^{13}$CS, and C$^{34}$S/$^{13}$CS, isotopologue line flux ratios with the elemental $^{32}$S/$^{34}$S$= 24.4\pm5.0$ in the vicinity of the Sun \citep{chin1996} and $^{12}$C/$^{13}$C$= 68\pm15$ in the local ISM \citep{milam2005,asplund2009,manfroid2009}. The CS/$^{13}$CS and CS/C$^{34}$S isotopologue ratios are found to be lower than the elemental ratios in 5/7 and 6/7 sources, respectively, from a factor of two up to one order of magnitude in I-04181, the source with the smallest disk-to-envelope ratio. The isotopologue ratios are only found to be compatible with elemental ratios in I-04016, for both of them, and in I-04302, for the CS/$^{13}$CS. As for the C$^{34}$S/$^{13}$CS ratios in all seven sources, they are all within a factor of two of the assumed elemental isotopic ratios, which is a very reasonable factor of uncertainty due to fractionation. Thus, it is very likely that both fainter isotopologues C$^{34}$S and $^{13}$CS are optically thin in our source sample. Hence, the fact that the brightest lines of CS appear in a different source than its fainter isotopologues is indeed likely due to optical depth effects.

\subsection{Resolved hyperfine structure}

\begin{figure}
    \centering
    \includegraphics[scale=1]{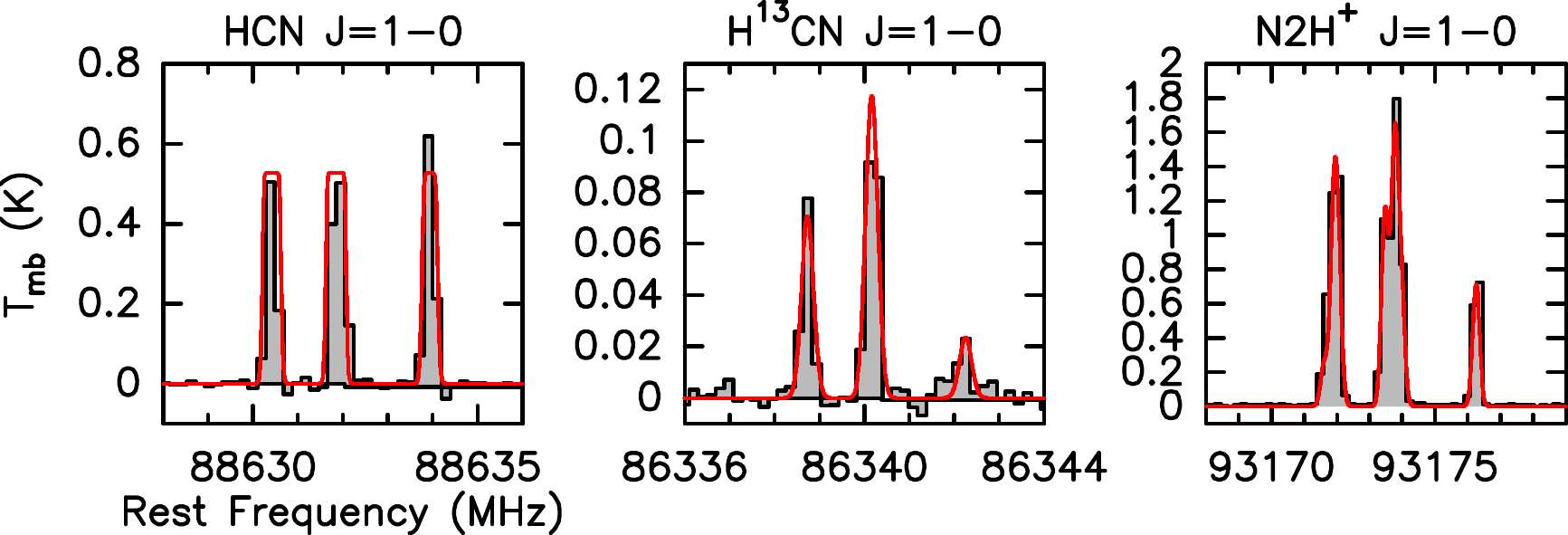}
    \caption{LTE modeling obtained for the hyperfine transitions of HCN, H$^{13}$CN, and N$_2$H$^+$  $J=1-0$ toward I-04166.}
    \label{fig:app-fit}
\end{figure}

HCN, H$^{13}$CN, and \ce{N2H+} lines split into a resolved set of hyperfine (hf) components, whose relative opacities are given by their Einstein coefficient for spontaneous decay and level degeneracy. In the LTE approximation, assuming that for each species all of the hf lines have the same excitation temperature, they can be used to infer the total opacity $\tau$ of the hf multiplet. Each hf component $k$ at rest frequency $\nu_k$ has an opacity $\tau_k=r_k\tau$ that scales accordingly with its relative intensity $r_k$. Assuming Gaussian profiles, the total opacity can thus be written as follows:
\begin{equation}
    \tau = \tau_0 \sum_k r_k \exp\left[-\frac{(v-v_k-v_0)^2}{2\sigma_v^2}\right],
\end{equation}
where the sum is over all hf components, $v_0$ is the systemic source velocity, and $v_k$ the velocity shift of a particular hf component $k$. Considering these assumptions, we derived the line opacity of the HCN, H$^{13}$CN, and \ce{N2H+} $1-0$ lines toward the sources of our sample where their hyperfine patterns are well-resolved (see Fig.~\ref{fig:30m_nitriles}), using LTE modeling through the \texttt{CASSIS} tool. The results are listed in Table~\ref{tab:opacities}  and shown for each molecule for one source, I-04166, in Fig.~\ref{fig:app-fit}, for illustration.
HCN is found to be highly optically thick in 4/7 sources, and marginally optically thick in the remaining three sources. \ce{H^13CN} and \ce{N2H+} are found to be optically thin in all sources where they are detected. 

\begin{table}
     \centering
     \caption{Opacity of the HCN, \ce{H^13CN}, and \ce{N2H+} $1-0$ lines toward our Class I source sample.}
     \begin{tabular}{lccccccc}
     \hline\hline
     Species & \multicolumn{7}{c}{Opacity}\\
     &\colhead{I-04302} & \colhead{I-04295} & \colhead{I-04365}& \colhead{I-04016}& \colhead{I-04166}& \colhead{I-04169}& \colhead{I-04181}\\
     \hline
       HCN  &  $\approx1$ & $\approx1$ & $\gg1$ &  $\approx1$ & $\gg1$ &$\gg1$  &$\gg1$\\
         \ce{H^13CN} & ...$^{(a)}$ & ... & $\ll1$ & $\ll1$ & $\ll1$  &$\ll1$ & $\ll1$\\
         \ce{N2H+} & ... & $\ll1$ & $\ll1$ & $\ll1$ & $\ll1$ & $\ll1$& $\ll1$\\
     \hline
     \end{tabular}
     \tablenotetext{a}{'...' indicates that the corresponding observations were insufficient (see Table~\ref{tab:intensities}) for that species and source.}
     \label{tab:opacities}
 \end{table}

\bibliography{manuscript}

\end{document}